\documentclass[useAMS,usenatbib]{mn2e}
\usepackage{times}
\usepackage{xkeyval}
\def\mpcoh{{\,h^{-1}\rm{ Mpc}}}

   \usepackage{wrapfig}
  \usepackage{gensymb}
 \usepackage[english]{babel}
  \usepackage{arydshln}
     \usepackage[pdftex]{graphicx}
     \usepackage[format=plain,justification=raggedright,singlelinecheck=false,font=small,labelfont=bf,labelsep=space]{caption}
\usepackage[justification=centering]{subcaption} 
     \usepackage{lipsum}
\usepackage{siunitx}
\usepackage[usenames,dvipsnames,svgnames,table]{xcolor}
\def\citejap#1{\citeauthor{#1}\ \citeyear{#1}}
\def\fig#1{Fig. \ref{#1}}
\def\tab#1{Table \ref{#1}}
\def\eq#1{Eq. \ref{#1}}
\usepackage{url}
\usepackage{sidecap}
\usepackage{hyperref}
\hypersetup{
  pdfauthor={...},
  pdftitle={...},
  pdfsubject={...},
  urlcolor=blue,
}

\title[GAMA: The galaxy luminosity function within the cosmic web]{Galaxy And Mass Assembly (GAMA): The galaxy luminosity function within the cosmic web}

\author[E. Eardley]{E. Eardley$^{1}$\thanks{E-mail:
ee@roe.ac.uk}, J.A. Peacock$^{1}$, T. McNaught-Roberts$^{2}$, C. Heymans$^1$, \newauthor P. Norberg$^2$, M. Alpaslan$^{3}$, I. Baldry$^4$, J. Bland-Hawthorn$^5$, S. Brough$^6$,\newauthor M.E. Cluver$^7$, S.P. Driver$^{10,12}$, D.J. Farrow$^{2,8}$, J. Liske$^{11}$, J. Loveday$^9$, \newauthor A.S.G. Robotham$^{12}$\\$^1$Institute for Astronomy, University of Edinburgh, Royal Observatory, Blackford Hill, Edinburgh EH9 3HJ, UK\\$^2$Institute for Computational Cosmology, Department of Physics, Durham University, South Road, Durham DH1 3LE, UK\\$^3$NASA Ames Research Centre, N232, Moffett Field, Mountain View, CA 94035, United States\\$^4$Astrophysics Research Institute, Liverpool John Moores University, IC2, Liverpool Science Park, 146 Brownlow Hill, Liverpool, L3 5R\\$^5$Sydney Institute for Astronomy, School of Physics A28, University of Sydney, NSW 2006, Australia\\$^6$Australian Astronomical Observatory, PO Box 915, North Ryde, NSW 1670, Australia\\$^7$Department of Physics, University of the Western Cape, Robert Sobukwe Road, Bellville, 7530, South Africa\\$^8$Max-Planck-Institut fuer extraterrestrische Physik, Postfach 1312, Giessenbachstr., 85748 Garching, Germany\\$^9$Astronomy Centre, University of Sussex, Falmer, Brighton BN1 9QH, UK\\$^{10}$School of Physics and Astronomy, University of St Andrews, North Haugh, St Andrews, KY16 9SS, UK\\$^{11}$European Southern Observatory, Karl-Schwarzschild-Str. 2, 85748 Garching, Germany\\$^{12}$ICRAR\thanks{International Centre for Radio Astronomy Research}, The University of Western Australia, 35 Stirling Highway, Crawley, WA 6009, Australia}

\begin{document}

\date{5$^{\rm th}$ November 2014} 


\maketitle


\begin{abstract}
We investigate the dependence of the galaxy luminosity function on geometric environment within the Galaxy And Mass Assembly (GAMA) survey. The tidal tensor prescription, based on the Hessian of the pseudo-gravitational potential, is used to classify the cosmic web and define the geometric environments: for a given smoothing scale, we classify every position of the surveyed region, $0.04<{z}<0.26$, as either a void, a sheet, a filament or a knot. We consider how to choose appropriate thresholds in the eigenvalues of the Hessian in order to partition the galaxies approximately evenly between environments. We find a significant variation in the luminosity function of galaxies between different geometric environments; the normalisation, characterised by $\phi^{*}$ in a Schechter function fit, increases by an order of magnitude from voids to knots. The turnover magnitude, characterised by $M^*$, brightens by approximately $0.5$ mag from voids to knots. However, we show that the observed modulation can be entirely attributed to the indirect local-density dependence. We therefore find no evidence of a direct influence of the cosmic web on the galaxy luminosity function.
\end{abstract}

\begin{keywords}
cosmology: observations, large-scale structure of Universe, surveys, galaxies: luminosity function
\end{keywords}
 \newpage
 \section{Introduction}

The galaxy luminosity function (LF) is central to studies of galaxy formation and evolution. A strong dependence on local environment of many galactic properties, such as morphology, star formation rate and colour, has long been established (e.g. \citejap{Dressler1980}, \citejap{Gomez2003}, \citejap{Balogh2004}). However, many models of galaxy formation assume only a very limited environmental impact. In standard halo-occupation models and some semi-empirical models, galaxy properties are assumed to depend only upon the mass of the host halo or its merger history (\citejap{Kauffmann1993}, \citejap{vdBosch2007}). With the existence of ever larger spectroscopic redshift surveys, such as SDSS 
and 2dFGRS, we are able to test these basic assumptions and search for evidence suggesting more complicated models. For example, a dependence of the galaxy LF on local density has been investigated and the LF has been shown to vary smoothly with overdensity, brightening continuously from void to cluster regions with no significant variation in the LF slope (e.g. \citejap{Croton2005}, \citejap{Tam14}
). \cite{Guo2014} measured the satellite LF of primary galaxies in SDSS and found a significant difference between galaxies residing in filaments and those that do not, suggesting that the filamentary environment has a direct effect on the efficiency of galaxy formation. 

There are many physical mechanisms that may be involved in determining the galaxy LF: mergers, tidal interactions and ram pressure gas stripping for example may all affect the luminosity of galaxies and induce an environmental dependence. Certainly some of these mechanisms must be influenced by the local matter density, purely through its impact on the population of dark-matter haloes -- which in turn affects the properties of the galaxies hosted by the haloes  (\citejap{Vale2004}, \citejap{Moster2010}). Much theoretical work concerning the formation, clustering and mass distribution of dark matter halos has already been undertaken. For example, the standard explanation for biased galaxy clustering uses the peak-background split formalism (\citejap{Bardeen1986}, \citejap{Cole1989}), in which the large-scale density field modulates the likelihood of collapse of haloes. But beyond this, it is conceivable that some galaxy properties may be linked not only with overdensity; for example, the tidal shear is also expected to affect the collapse of haloes, with inevitable knock-on effects on galaxy properties (\citejap{Sheth2001}).

With the advent of numerical simulations we are able to test in more detail the extent to which different properties of the environment may influence LSS formation. \cite{Hahn2009} find the mass assembly history of halos to be influenced by tidal effects, and note that tidal suppression of small halos may be especially effective in filamentary regions. \cite{Ludlow2011} used cosmological $\Lambda$CDM simulations to test the central ansatz of the peaks formalism, in which halos evolve from peaks in the linear density field when smoothed with a filter related to the halos characteristic mass. Although they found the majority of halos to be consistent with this picture, they identify a small but significant population of halos showing disparity and find these halos are, on average, more strongly compressed by tidal forces.

The visible manifestation of such tidal forces is the striking way in which gravitational instability rearranges the nearly homogeneous initial density field into the cosmic web. Numerical simulations and large galaxy surveys both show an intricate filamentary network of matter: large, underdense void regions are surrounded by 2-dimensional sheets and 1-dimensional filamentary structures, which meet to form highly overdense nodes, or knot regions, where many halos reside. We shall use the term `geometric environment' to denote these different regions of the cosmic web. Recent years have seen an increased interest in methods of classifying the cosmic web (see e.g. \citejap{Cautun2013} for an overview). Many of these studies have been applied to numerical simulations, finding some promising detection of LSS alignments with filaments (\citejap{Codis2012}, \citejap{Forero-Romero2014})
.  Studies of geometric environments in observational datasets have more often focused on identifying individual structures such as voids or filaments rather than on classifying the global volume. In this work we present an application of the tidal tensor prescription, based on the second derivatives of the gravitational potential, to the Galaxy And Mass Assembly (GAMA) spectroscopic redshift survey (\citejap{Driver2011}, Liske et al. submitted). We classify the surveyed volume as either a void, a sheet, a filament or a knot by approximating the dimensionality of collapse.
In this way, we are able to calculate a conditional luminosity function as a function of location within the cosmic web. Our motivation is to search for any correlation of galaxy properties with this non-local aspect of the density field. Of course, some galaxy properties may be affected in a completely local manner (see e.g. \citejap{Wijesinghe2012}, \citejap{Brough2013} and \citejap{Robotham2013} for previous studies of the dependence of GAMA galaxy properties on local environments), so in parallel we will need to track the dependences that are purely functions of overdensity.
  
This paper is structured as follows: In section \ref{WEB} we present the method used for the environmental classifications and discuss some of the technical issues and limitations of applying this method to observational datasets. The data sample and resulting environments are presented in section \ref{GAMA_section}. In section \ref{LF_section} we present the conditional LF and test the direct influence of the web by comparing our measurement with LFs measured for galaxies with matching local density distributions. Finally, in section \ref{summary} we discuss and summarise our results.
 
 We adopt a standard ${\rm \Lambda}$CDM cosmology with $\Omega_{\rm M}=0.25$, $\Omega_{\Lambda}=0.75$ and $H_{0}=100\,h\,{\rm kms}^{-1}{\rm Mpc}^{-1}$, and note that apart from the gridding process, where galaxies are assigned to a Cartesian grid, the classification of geometric environments implemented in this work is cosmology independent.

\section{Classifying the cosmic web}
 \label{WEB}
Although the cosmic web is clearly visible in all sufficiently detailed observed and simulated distributions of matter, its complexity and variety of scales, shapes, densities and dimensionality makes it nontrivial to quantify. A number of different approaches have been proposed and developed: minimal spanning tree methods have been used to detect filaments (\citejap{Barrow1985}, \citejap{Alpaslan2014}); topological methods based on Morse theory (\citejap{Sousbie2011}) and morphological methods based on feature extraction techniques (\citejap{Aragon-Calvo2010}) and the watershed transform (\citejap{Platen2007}) have all been used to identify the full range of web components. Additionally, both the tidal tensor and the velocity shear tensor, with theoretical motivations drawn from Zel'dovich theory (\citejap{Zeldovich1970}), are able to produce good visual matches to the cosmic web (\citejap{Forero-Romero2009}, \citejap{Hoffman2012}). Similarly, the ORIGAMI method of structure identification (\citejap{Falck2012}), which considers the folding of a 3D manifold in 6D phase space, has been successfully applied to simulations. However, many of these methods cannot be applied to observational data as they require information on the peculiar velocity of galaxies. Though each method has its advantages, we choose to follow the approach of \cite{Hahn2007} based on the tidal tensor prescription, for its applicability to both simulated and observational datasets, and for its appealing theoretical underpinnings  (see \citejap{Alonso2014} for a discussion of Gaussian statistics and the theoretical conditional halo mass function in this definition of the web).

The tidal tensor prescription is in essence a stability criterion based on linear dynamics at each point in space.
Each location is classified as a void, a sheet, a filament or a knot depending on whether structure is said to be collapsing in 0, 1, 2 or 3 dimensions respectively. This can be derived from knowledge of the gravitational potential field, $\Phi$, using the tidal tensor, $T_{ij}$, defined as the matrix of second derivatives of $\Phi$:
\begin{equation} 
\label{TT}
T_{ij}= \frac{\partial^{2}\Phi}{\partial{r_{i}}r_{j}}.
\end{equation}
The three real eigenvalues of the symmetric $T_{ij}$ allow us to make our classification; the number of positive eigenvalues is equivalent to the dimension of the stable manifold at the point in question.

\subsection{Measuring the tidal tensor}
\label{NI}

To calculate $T_{ij}$ we first require the matter overdensity field, $\delta$. Lacking direct knowledge of the underlying dark matter, we work with the pseudo-gravitational potential that is sourced by the number density of galaxies. The uncertainties introduced through using galaxies -- biased tracers in redshift space -- to estimate the real-space density field are discussed in appendix \ref{rsdsec}. There we show an analysis of simulated data which indicates that using galaxies to estimate the underlying density field changes the classifications for $\textless{20}\%$ of the volume. 

Galaxies are assigned to a Cartesian grid with cubic cells of width $R_{c}=3\mpcoh$ by cloud-in-cell interpolation, which uses multilinear interpolation to the eight nearest grid points to each galaxy. Experimentation with the value of $R_{c}$ has shown that results converge by  $R_{c}\approx3\mpcoh$ and any further variation caused by using smaller grid cells is negligible. The overdensity of each cell is given by

\begin{equation} 
\label{odeneq}
{\delta}= \frac{{N_{\rm obs}}}{{N_{\rm R}}} - 1,
\end{equation}
where ${N_{\rm obs}}$ is the number of observed galaxies within the cell after the interpolation, and $N_{R}$ is an estimate of the corresponding number that would have been observed if there were no clustering. More specifically, $nN_{\rm R}$ is the interpolated number density of a random catalogue generated by cloning real GAMA galaxies in our sample $n$ times (we use $n=400$) and distributing them randomly within the maximum volume over which they can be observed (Farrow et al. in prep., \citejap{Cole2011}). 

In order for the tidal tensor to be well defined, the discrete density field must be smoothed. The purpose of this step is to suppress shot noise, and also to remove extreme non-linearities. We smooth the density field with a Gaussian filter of width $\sigma_{s}$. The cloud in cell interpolation also inevitably introduces additional smoothing. By Taylor expanding the Fourier space window function for cloud-in-cell interpolation, one can show that this additional smoothing is approximately equivalent to smoothing with a Gaussian of width $\sigma_{c}=R_{c}/\sqrt{6}$. Hence, the effective smoothing scale is $\sigma=\sqrt{\sigma_{c}^{2}+\sigma_{s}^{2}}$, and can be thought of as the typical length scale on which we are determining dynamical stability. In the spirit of the Zel'dovich approximation, we should filter until we reach scales where only a moderate amount of shell-crossing has occurred, linking the observed density field to the initial conditions. With this, and the number density and survey geometry of GAMA in mind, we chose to use effective smoothing scales of $\sigma=4$ and $10\mpcoh$ (in order to show how the results depend on resolution near the nonlinear scale).

An immediate practical problem is how to deal with the survey boundaries during this smoothing process given that we do not have knowledge of the density field beyond the surveyed region. Zero-padding the survey, by setting $\delta=0$ for regions outside of the survey boundaries, would bias the density field inside the survey. In order to ameliorate this problem, before the smoothing process we populate the volume outside of the survey with cloned galaxies `reflected' along the boundaries of the field, which is approximately equivalent to using a weighted smoothing kernel. This method of reflecting cloned galaxies is discussed in more detail in appendix \ref{ee}.

The pseudo-gravitational potential field and its second spatial derivatives can be derived from the smoothed galaxy-overdensity field, $\delta$, by working in Fourier space. 
The potential, $\Phi$, can be obtained by solving Poisson's equation
\begin{equation} 
\label{Poisson}
\nabla^{2}{\Phi}= 4{\pi}G\bar{\rho}\delta=\alpha+\beta+\gamma,
\end{equation}
where $\bar{\rho}$ 
is the average matter density of the Universe, 
$G$ the gravitational constant and  $\alpha,\beta,\gamma$ are the 3 eigenvalues of the diagonalized Hessian of $\Phi$. However, it is useful to consider the dimensionless potential, $\tilde{\Phi}$, and the dimensionless eigenvalues ${\lambda_{1}}, {\lambda_{2}}$ and ${\lambda_{3}}$, found by factoring  $4{\pi}G\bar{\rho}$ out of \eq{Poisson}:

\begin{equation} 
\nabla^{2}\tilde{\Phi}= \delta={\lambda_{1}}+{\lambda_{2}}+{\lambda_{3}}.
\label{PoissonScaled}
\end{equation}
We note that with this normalisation the pseudo-gravitational potential of \eq{PoissonScaled} is independent of bias in the limit of linear bias.

In Fourier space the dimensionless potential and its Hessian, the tidal tensor, $T_{ij}$, are given by
\begin{equation} 
\tilde{\Phi}_{k}=-\frac{\delta_{k}}{k^{2}}\hspace{5mm}{\rm{and}}\hspace{5mm} \tilde{T_{ij}}=\frac{\partial^{2}\tilde\Phi_{k}}{\partial_{i}\partial_{j}}= \frac{k_{i}k_{j}\delta_{k}}{k^{2}},
 \end{equation}
 with  $k=\sqrt{(k_{i}^{2}+k_{j}^{2}+k_{k}^{2})}$.

The eigenvalues of $T_{ij}$ are calculated for each cell of the Cartesian grid and comparison with an eigenvalue threshold, discussed below, leads to the classification of the region within the cell.

\subsection{The eigenvalue threshold}
\label{eigchap}
A positive but infinitesimally small eigenvalue implies that structure is collapsing along the corresponding eigenvector, but it may not reach non-linear collapse for a significant period of time, if ever. This leads to an overestimation of the number of collapsed dimensions and the resulting classifications are a poor match to the visual impression of the web. Hence, in order to account for the finite time of collapse, we follow the extension of \cite{Forero-Romero2009} and introduce an eigenvalue threshold, $\lambda_{\rm th}$, as a free parameter of the tidal tensor prescription method of classifying geometric environments. We use the number of eigenvalues greater than this threshold to define our environments rather than the number greater than zero. After the normalisation discussed in section \ref{NI}, \eq{PoissonScaled} shows that the sum of the eigenvalues will be equal to the density contrast, hence we expect an appropriate threshold parameter will be of order unity. 

With the introduction of  $\lambda_{\rm th}$, the 3 eigenvalues calculated for each location lead us to classify regions as follows (with $\lambda_{3}<\lambda_{2}<\lambda_{1}$):
\newline
    \begin{itemize}
      \item {\bf Voids}: all eigenvalues below the threshold \\
                \hspace{2cm}  ($\lambda_1<\lambda_{\rm th}$).
      \item {\bf Sheets}:  one eigenvalue above the threshold \\
           \hspace{1cm} ($\lambda_1>\lambda_{\rm th}$, $\lambda_2<\lambda_{\rm th}$).
      \item {\bf Filaments}: two eigenvalues above the threshold \\
            ($\lambda_2>\lambda_{\rm th}$, $\lambda_3<\lambda_{\rm th}$).
      \item {\bf Knots}: all eigenvalues above the threshold \\
            ($\lambda_3>\lambda_{\rm th}$).
    \end{itemize}
    
\begin{figure}
\includegraphics[width=0.48\textwidth]{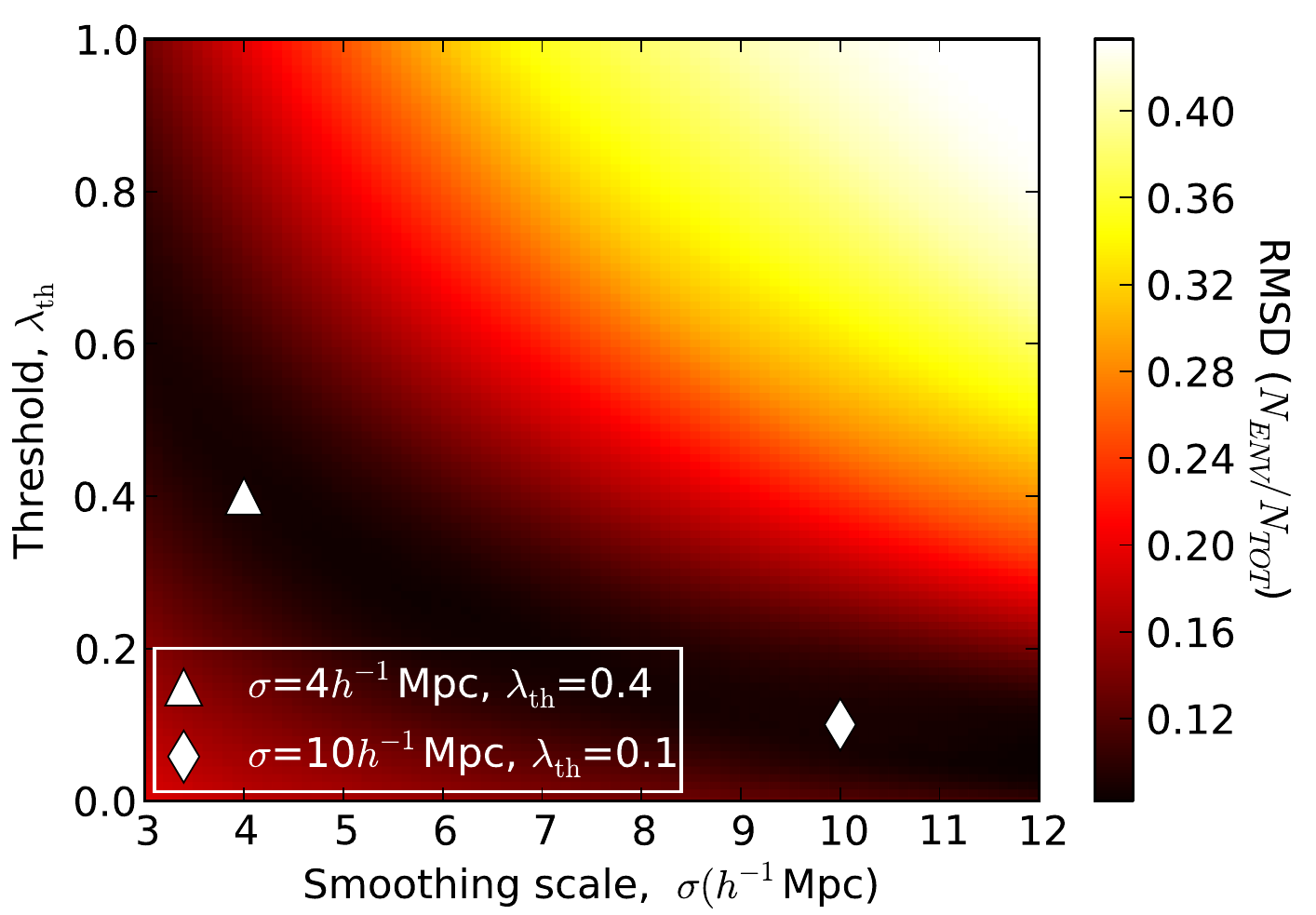}
\caption{We wish to choose our two free parameters, the eigenvalue threshold, $\lambda_{\rm th}$, and the smoothing scale, $\sigma$, in a way that optimises the resulting statistics by assigning a comparable number of objects to each geometric environment. This plot displays the root mean squared dispersion (RMSD, as defined by Eq. \ref{rmsdeq}), in the number of galaxies in our sample which are assigned to each geometric environment as a function of $\lambda_{\rm th}$ and $\sigma$ used to generate the classifications. The dark curve represents the statistically optimal region in the parameter space, motivating our choice of two parameter sets: ($\sigma$,$\lambda_{\rm th}$) = (4$\mpcoh$, 0.4) and  (10$\mpcoh$, 0.1), as indicated in the figure.}
\label{rmsdnewdots}
\end{figure}

In this paper we present results for two smoothing scales, $\sigma=4$ and 1$0\mpcoh$, chosen to study a wide range of scales whilst reflecting the limitations caused by the number density and survey volume of GAMA. The choice of  $\lambda_{\rm th}$ is similarly arbitrary; whilst it changes the classification of the web, there is no strict definition of what constitutes a void region for example, and hence our classifications can be adapted to suit the task at hand. One could use the spherical collapse model to explicitly derive the eigenvalue threshold which corresponds to collapse along the eigenvector by equating the collapse time with the age of the Universe, but the invalid assumption of spherical isotropic collapse would allow for only a rough estimate of the true threshold. An alternative approach is to choose the threshold that produces the best visual agreement of the resulting web with the distribution of matter, but such subjectivity is undesirable. Instead, in this work we choose to set the value of the eigenvalue threshold in order to optimise the statistical significance of any measurement that we might choose to make in the different environments, i.e. to allocate the objects under study to the four environments as equally as possible.
To do so, for a variety of parameter sets we calculate the root mean square dispersion (RMSD) of the fraction of all galaxies in the selected sample (see section \ref{samplesec}) classified as each of the four geometric environments from the mean fraction. That is, we calculate the RMSD, defined as

\begin{equation}
{\rm RMSD}(X_{i})=\sqrt{\frac{\sum_{0}^{3}(X_{\rm i}-0.25)^{2}}{4}}\qquad X_{i}=\frac{N_{\rm ENV, i}}{N_{\rm TOT}}
\label{rmsdeq}
\end{equation}
where $N_{{\rm ENV}, i}$ is the number of galaxies belonging to environment $i$, and $N_{\rm TOT}$ is the total number of galaxies in the full sample, so that environment $i$ holds a fraction $X_{i}$ of all galaxies. \fig{rmsdnewdots} shows this root mean squared dispersion in environmental number count as a function of the smoothing scale, $\sigma$, and the imposed eigenvalue threshold, $\lambda_{\rm th}$. 
We wish to minimise this quantity in order to ensure that all environments hold enough galaxies to maintain a low level of statistical uncertainty, which is essential in order to look for potentially small modulations due to geometric environments. No choice of parameters produces an exactly equal split, where each environment holds $25\%$ of galaxies, but there exists a range of parameters such that each environment holds at least $10\%$. The dark shaded region represents this optimal parameter space --  for smaller smoothing scales we require a higher threshold in order to maintain a near-comparable split of galaxies and vice versa. Based on this, we focus on environments defined by the parameter sets shown by the symbols in the figure: ($\sigma$,$\lambda_{\rm th}$) = (4$\mpcoh$, 0.4) and  (10$\mpcoh$, 0.1). The resulting partition of galaxies and of the survey volume between the environments defined by these two parameter sets are given in \tab{Schtab}. We note that these parameters do in fact produce environments that seem visually plausible, even though this was not a criterion.

 \section{Application to GAMA}
\label{GAMA_section}

\label{samplesec}
\subsection{Galaxy And Mass Assembly}
\label{samplesec}

We use data from the Galaxy and Mass Assembly (GAMA) survey (\citejap{Driver2009}, \citejap{Driver2011}, Liske et al. submitted), a spectroscopic redshift survey 
 split between 5 regions. GAMA is an intermediate redshift survey, bridging the gap between wide-field surveys such as 2dFRGS (\citejap{Colless2001}) and SDSS (\citejap{York2000}) and high-redshift deep field surveys such as VIPERS (\citejap{Garilli2014}). By surveying a cosmologically representative volume whilst maintaining an impressive $>98\%$ redshift completeness in the equatorial fields (\citejap{Robotham2010b}), GAMA provides an ideal dataset with which to study the modulation of galactic properties by large-scale environments. In full, GAMA observes galaxies out to $z\simeq0.5$ and $r<19.8$; but in this work we study the lower redshift regime where the number density and magnitude range of observed galaxies is statistically sufficient. 
For consistency with the previous analysis of the environmental dependence of the luminosity function within GAMA by \cite{Tam14} (hereafter MNR14) we use a sample of $113000$ galaxies satisfying $0.04<z<0.263$ selected from the 3 equatorial regions of GAMA: G09, G12 and G15, each spanning  $12\degree\times5\degree$. When testing the effects of the chosen sample, we found no benefit to restricting the catalogue to a volume-limited subset, hence no absolute magnitude cuts are imposed. We use all galaxies with a GAMA redshift quality rating of ${\tt nQ}>2$, indicating the redshift is sufficiently reliable to be included in scientific analyses, and an appropriate visual classification flag $({\tt VIS\_CLASS} = 0, 1$ or $255$; \citejap{Baldry2010}). 

\begin{figure*}
	\makebox[1\linewidth][c]{%
\hspace{0.15cm}
		\begin{subfigure}{0.2\textwidth}
			\includegraphics[scale=1.4]{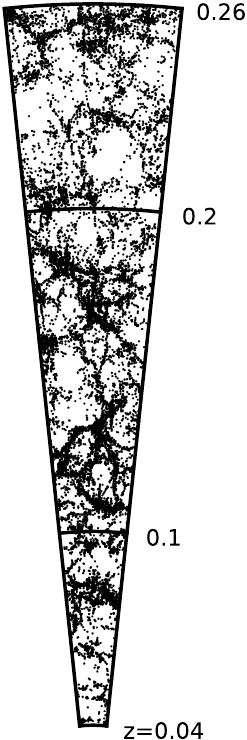}
			\caption{}
			\label{fig:galdist}
		\end{subfigure}
\hspace{0.1cm}
   		 \begin{subfigure}{0.2\textwidth}
			\includegraphics[scale=1.36]{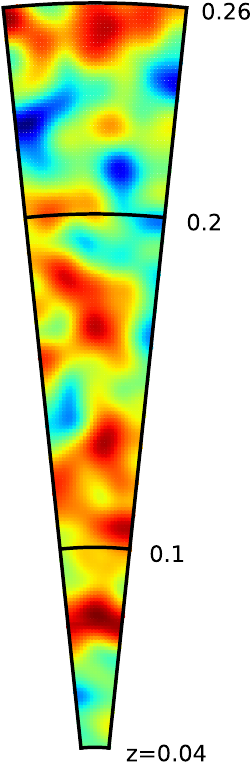}
			\caption{}
			\label{fig:galdencon}
		\end{subfigure}

		\begin{subfigure}{0.2\textwidth}
			\includegraphics[scale=1.0]{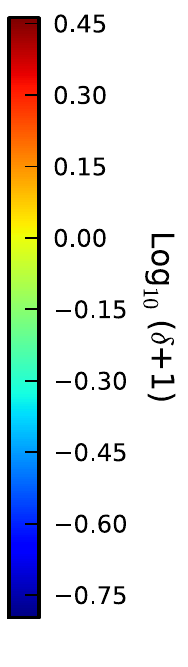}
			\label{fig:galdenconbar}
		\end{subfigure}
\hspace{-1.4cm}

		\begin{subfigure}{0.2\textwidth}
			\includegraphics[scale=1.4]{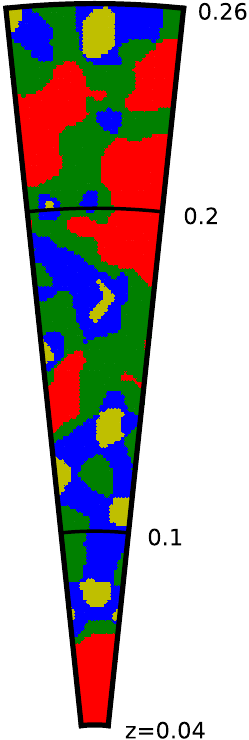}
			\caption*{\bf{(c)} $\sigma=10\mpcoh$}
			\label{fig:galenv1}
		\end{subfigure}

				\begin{subfigure}{0.05\textwidth}
\vspace{1cm}
\hspace{-1.345cm}
			\includegraphics[scale=1]{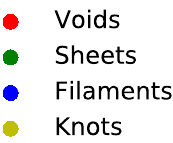}

			\label{fig:galenv1}
		\end{subfigure}	
		\hspace{-0.5cm}
		\begin{subfigure}{0.2\textwidth}

\hspace{0.5cm}

			\includegraphics[scale=1.4]{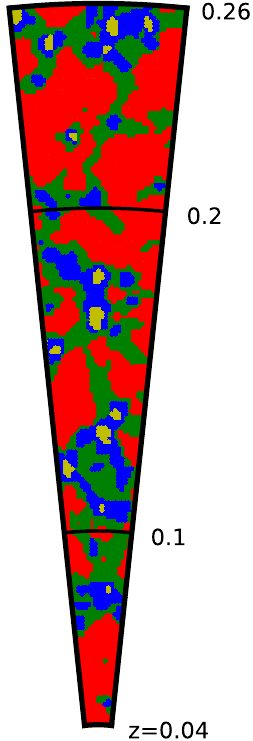}
			\caption*{\bf{(d)} $\sigma=4\mpcoh$}
			\label{fig:galenv1}

		\end{subfigure}		
		        
	}
	\caption{An example of the classification of geometric environments within the GAMA G12 field. {\bf(a)} Distribution of galaxies within $\pm\ang{1}$ of the central declination. {\bf(b)} The density contrast field, $\delta$, derived from {\bf(a)} after interpolation of the galaxies on to a Cartesian mesh and smoothing with a Gaussian kernel of effective width $\sigma=10\mpcoh$, with a colour scale proportional to $\log_{10}(\delta + 1)$ as given by the colour bar to the right. {\bf(c)} The resulting geometric environment classifications, with an eigenvalue threshold of $\lambda_{\rm th}=0.1$, from the smoothed density contrast field in {\bf(b)}. {\bf(d)} The geometric environments for the second parameter set, $(\sigma, \lambda_{\rm th})=(4\mpcoh, 0.4)$. Environments are colour coded as shown in the key, e.g.: red, green, blue and yellow for voids, sheets, filaments and knots respectively. Whilst panel {\bf(a)} shows a 2D projection of galaxies, the slices shown in panels {\bf(b)}, {\bf(c)} and {\bf(d)} show the 2D plane of the central declination; they show the value (density contrast or environment) of whichever cell is intersected by the central declination.}
\label{G12conesden}
\end{figure*}

\label{galenvs}
\begin{figure*}
\makebox[1.\linewidth][c]{%
\hspace{-1cm}
    \begin{subfigure}{0.2\textwidth}
        \includegraphics[scale=1.6]{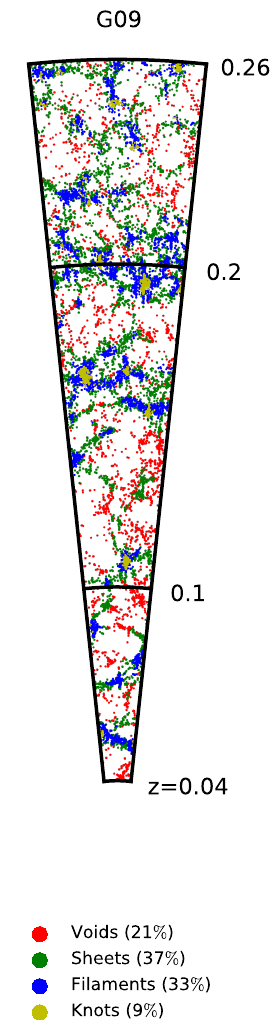}
        \label{fig:galdist}
   \end{subfigure}

\hspace{2.7cm}
    \begin{subfigure}{0.2\textwidth}
\includegraphics[scale=1.6]{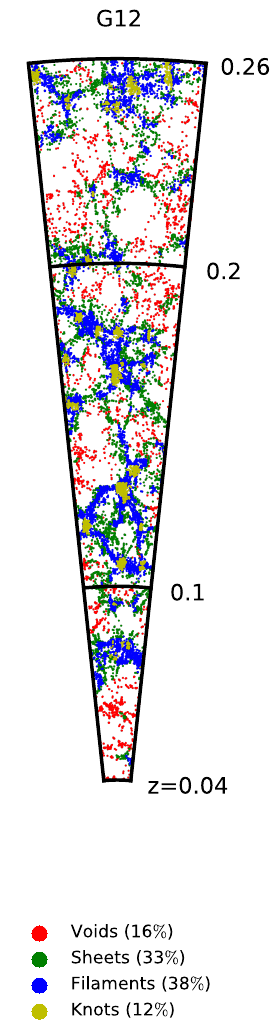}
 \label{fig:galdencon}
\end{subfigure}
      \hspace{2.3cm}
    \begin{subfigure}{0.2\textwidth}
  \hspace{2.3cm}
        \includegraphics[scale=1.6]{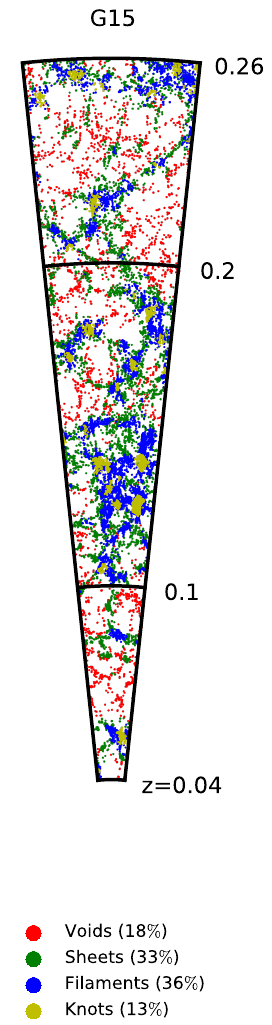}
        \label{fig:galenv1}
    \end{subfigure}
 }%
    \caption{The distribution of galaxies in the 3 equatorial GAMA fields within $\pm\ang{1}$ of the central declination. Galaxies are colour coded by their resulting geometric environment classification after smoothing with a Gaussian of width $\sigma=4\mpcoh$ and applying a threshold of $\lambda_{\rm th}=0.4$. For each of the GAMA fields, the percentage of galaxies within each of the four environments is shown in the keys beneath the cones.}
    \label{fig:4galcones}
\end{figure*}
\begin{figure*}
	\makebox[1\linewidth][c]{%
		\begin{subfigure}{0.5\textwidth}
			\includegraphics[width=1\textwidth, scale=0.1]{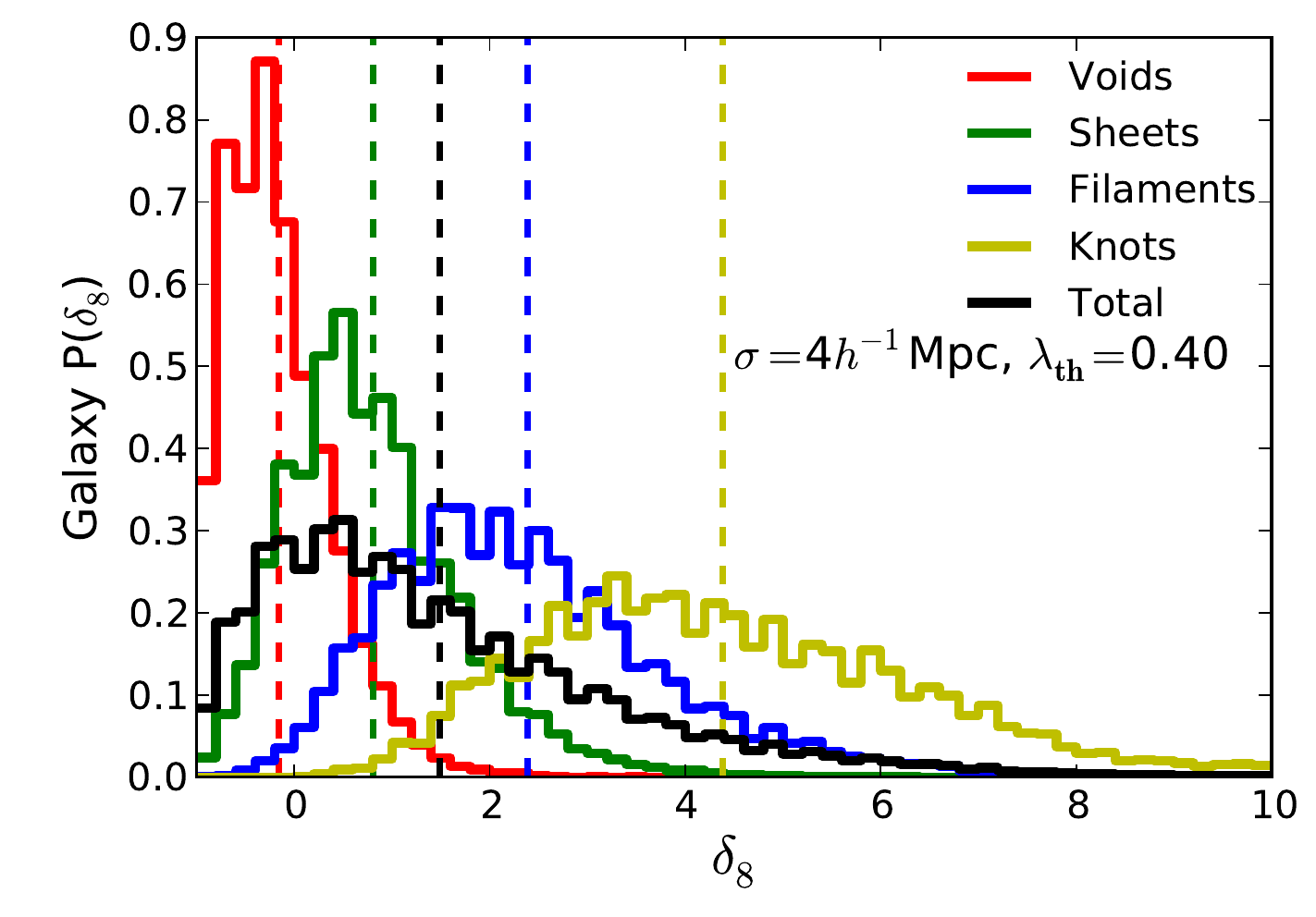}

		\end{subfigure}
    		\begin{subfigure}{0.5\textwidth}

			\includegraphics[width=1\textwidth, scale=0.1]{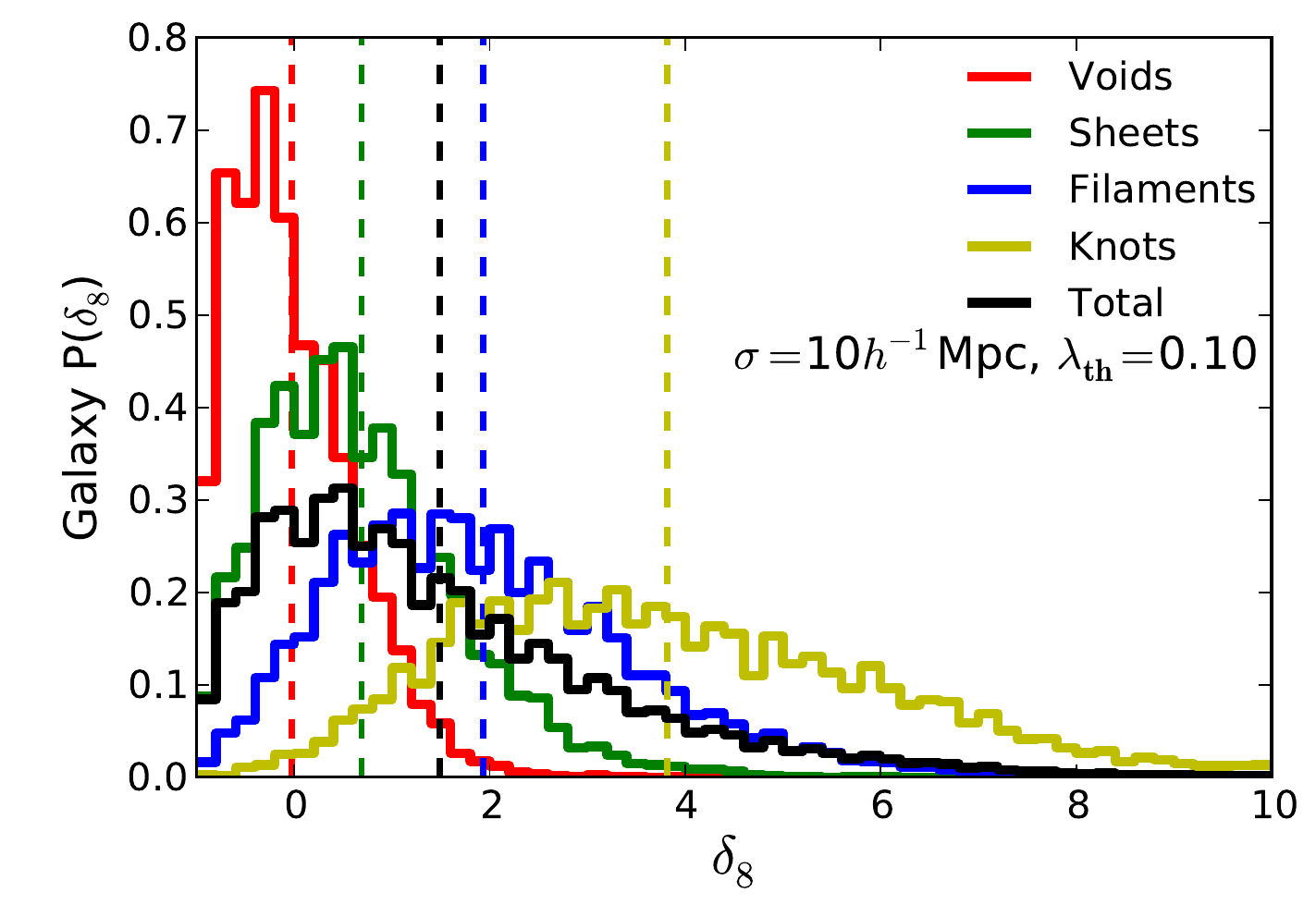}
		\end{subfigure}
	}	
	\caption{Distribution of local overdensities for galaxies split by geometric environment. Dashed lines indicate the average overdensities, as given in \tab{Schtab}, of all galaxies within each environment. The overdensity, $\delta_{8}$, is derived from the number of galaxies within a sphere of radius 8$\mpcoh$. The overdensity increases as the dimension of the environment reduces; but because there is a wide range of overdensities in each environment we can look for a dependence of galaxy properties on $\delta_{8}$ and geometric environment separately.} 
		
\label{hists}
\end{figure*}

\subsection{Observed cosmic web within GAMA}
We construct a density field from the GAMA galaxies detailed above for each of the three equatorial fields separately. As discussed in section \ref{NI}, and in more detail in appendix \ref{ee}, the volume immediately outside the survey region is populated with cloned galaxies reflected along the boundaries of each field. This is in order to reduce the effects of the survey geometry when smoothing the density field. 

\fig{G12conesden} illustrates the main steps in the classification of one of the GAMA fields. The galaxies are interpolated onto a Cartesian grid and an overdensity field generated by comparison with the catalogue of randomly positioned cloned galaxies. This overdensity field is smoothed by a Gaussian window function of width $\sigma_s$ in Fourier space. Following that, the potential and its second derivatives are calculated, from which the 3 eigenvalues can be derived for each cell of the grid. We approximate the dimensionality of collapse by the number of eigenvalues above the chosen eigenvalue threshold and from this each cell is classified as either a void, sheet, filament or knot. Finally, the galaxy catalogue is split into four environmentally defined subcatalogues by assigning the galaxies the geometric environment of the cell in which they reside. 

In \fig{G12conesden}a we plot the distribution of all galaxies within $\pm\ang{1}$ of the central declination of the G12 field. \fig{G12conesden}b is the resulting density field along the central declination of the G12 field, after smoothing with an effective smoothing scale of $\sigma=10\mpcoh$.  In \fig{G12conesden}c and \fig{G12conesden}d we show the resulting geometric environments of the central declination slice of the G12 field for our two parameter sets, $(\sigma, \lambda_{\rm th}) = (10\mpcoh, 0.1)$ and $(4\mpcoh, 0.4)$ respectively. As may be expected, both figures display a similar basic skeleton, with the larger smoothing scale resulting in larger geometric structures. The knots in particular are visibly larger in \fig{G12conesden}c.

\subsection{Geometric environments of GAMA galaxies}
The geometric environments of all galaxies in the sample are defined by the classification of the cell they belong to. In \fig{fig:4galcones} we plot the geometric environment classifications of galaxies around the central declination slice of each of the 3 GAMA fields. We show here environments defined for the parameter set $(\sigma, \lambda_{\rm th})=(4\mpcoh, 0.4)$. Although the sheets are not visually well captured in a 2D figure, the galaxies in filaments stand out clearly, particularly when the filament happens to lie in the plane of the figure \footnote{For an animated view of the 3D distribution of geometric environments, see \url{http://www.roe.ac.uk/~ee/GAMA}}. The void galaxies are in less populated regions but sometimes exhibit small-scale clustering which has been smoothed out during the filtering process. 

As well as having a distinct shape, the different geometric environments are also strongly distinct in terms of density. The distributions of local overdensities within each geometric environment are shown in \fig{hists}, where the overdensity is calculated from the number of galaxies within a sphere of radius 8$\mpcoh$ centred on the location of each galaxy rather than the grid-based overdensity measure used during the environment classification process. This density measure follows from the work of \cite{Croton2005} and is chosen for consistency with MNR14; it involves selecting a density-defining-population of galaxies which form a volume-limited sample over our redshift range. All galaxies within the $8\mpcoh$ radius contribute to the density measure if they are part of the density-defining-population, including the galaxy for which we are measuring the density. We convert the measured densities to $\delta_8$, our measure of overdensity, by comparison with the effective mean density within the sample.

As expected, the average overdensity increases as the dimensionality of the environment decreases (note that the 3D voids are the highest dimension of environment, with knots considered to have the lowest dimensionality, having collapsed in all dimensions); most void galaxies are found in underdense regions, almost no knot galaxies reside in underdense regions and instead live in highly overdense areas. One may be surprised that a significant proportion of voids are found to be slightly overdense; a similar result was found in the analysis of simulated data by \citejap{Alonso2014}. If we retain the simplicity of an environment classification using the tidal tensor, this feature cannot be entirely removed. The fraction of overdense voids is reduced if the threshold, $\lambda_{\rm th}$, is made smaller, but extreme low thresholds do not produce a good visual impression of the web and result in apparently 3D regions being classified as a 2D sheet.
 The broad distribution of densities within each environment shows the geometric environment holds more information than the density alone. Environments derived from the $10\mpcoh$ smoothed density field show a slightly larger spread of densities in any given environment than for the $4\mpcoh$ field, though the distributions are relatively similar.

An alternative method of classifying LSS within GAMA using minimal spanning trees was presented in \cite{Alpaslan2014}. In appendix \ref{mehmet} we compare our results, finding a strong visual agreement for the filamentary regions but limited agreement in our classification of voids. With the somewhat flexible definitions of geometric environments this is neither a surprise nor cause for concern. On the contrary, it illustrates the variety of meanings of terms such as voids, even within the context of the cosmic web. Hence, when interpreting any results in the context of the web it is important for one to have a clear quantitative understanding of the how the environments in question are defined.

 \section{Luminosity functions and geometric environment}
\label{LF_section}

The galaxy LF is measured independently for each geometric environment, using k-corrected and luminosity evolution corrected absolute $r$-band magnitudes, $M_{e}^{r}$,  and following the approach taken in MNR14. This method adopts the step-wise maximum likelihood estimator (\citejap{Efstathiou1988}), and normalizes the LF taking into account the effective fraction of the volume covered by a given geometric environment ($f_{{\rm env}}$, estimated by counting the number of galaxies in the unclustered random catalogue that fall within the regions classified as each environment):
\begin{equation} 
N_{\rm env} =  f_{\rm env} \Omega_{\rm tot} \int_{z_{\rm 1}}^{z_{\rm 2}}\, {\rm d}z \frac{{\rm d}V}{{\rm d}z{\rm d\Omega} }\int_{M_{{\rm faint}(z)}}^{M_{{\rm bright}(z)}}  \phi(M)\,{\rm d}M
\end{equation}
for a galaxy sample with $N_{\rm env}$ galaxies in a given environment, redshift limits $z_{\rm 1}$ and $z_{\rm 2}$ and total solid angle ${\rm \Omega_{\rm tot}}$.

\begin{figure}
	\begin{subfigure}{0.5\textwidth}
		\includegraphics[scale=0.44]{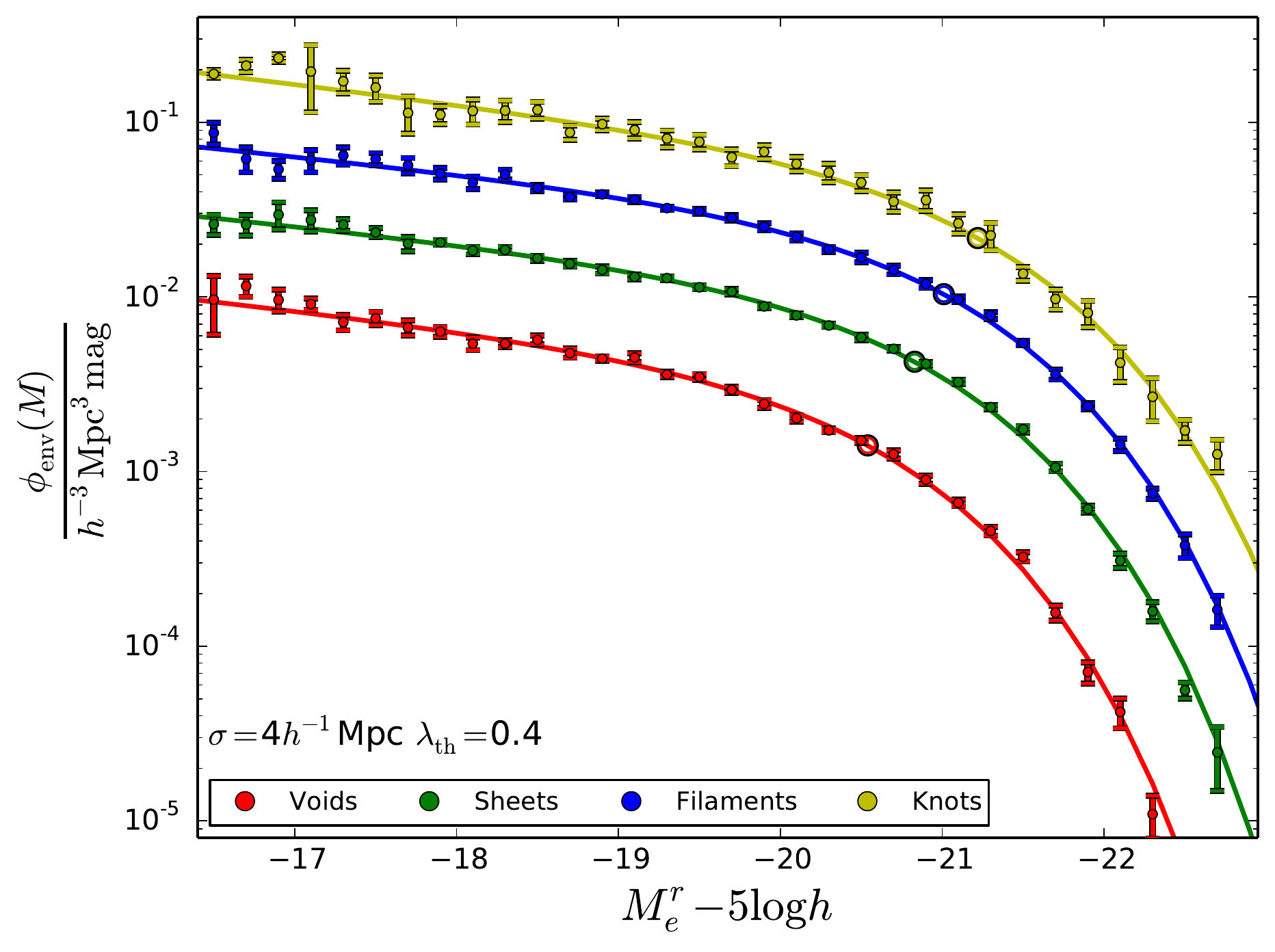}
	\end{subfigure}

	\begin{subfigure}{0.5\textwidth}

		\includegraphics[scale=0.44]{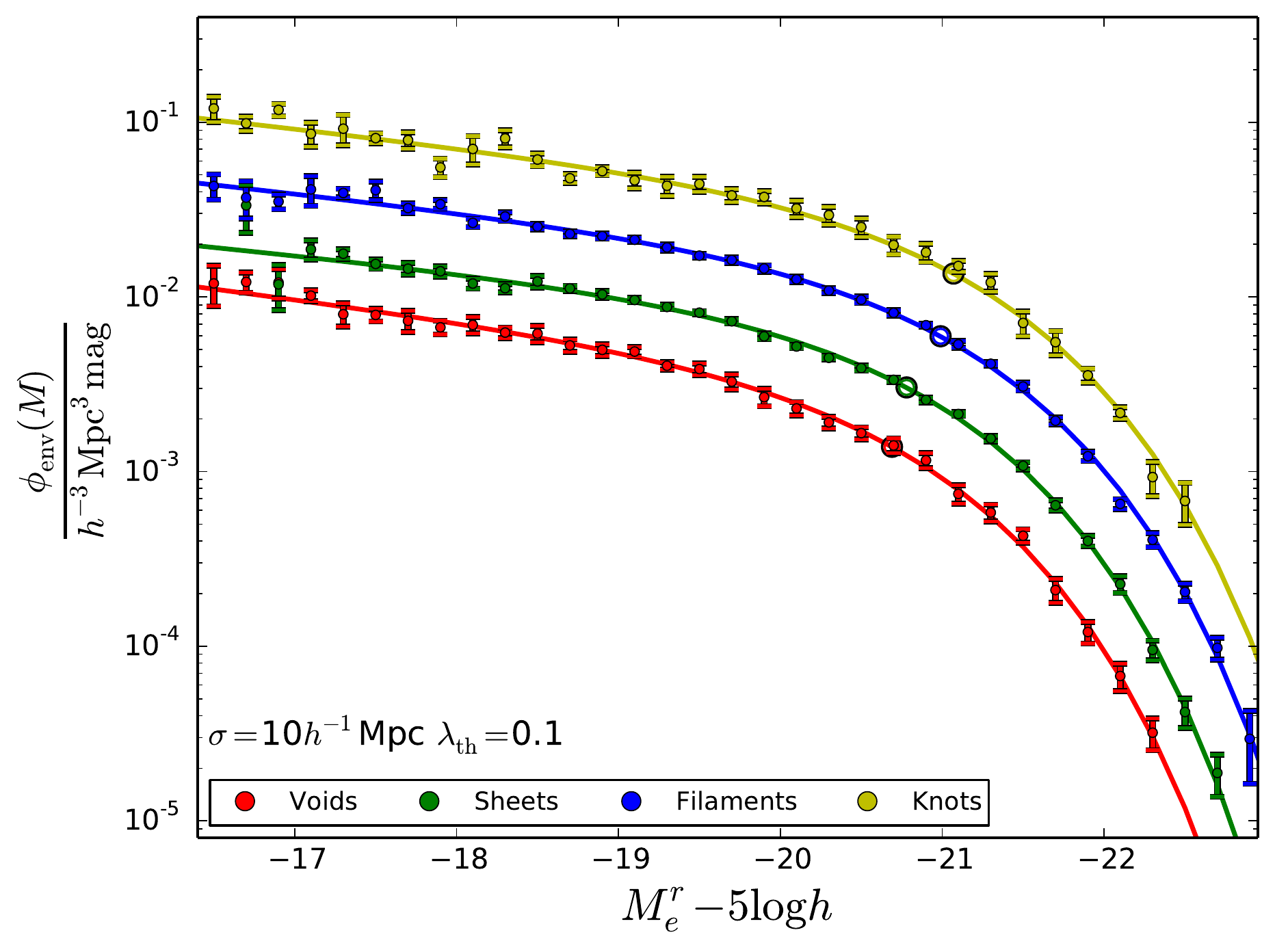}

	\end{subfigure}
	\caption{The galaxy luminosity functions and corresponding jackknife errors of the 4 subcatalogues produced by splitting the GAMA sample according to geometric environment, defined with $\sigma=4$ or $10\mpcoh$. Solid lines show best fitting Schechter functions for each conditional LF, open circles show with the best fitting value of $M^*$, given in \tab{Schtab}. The normalisation, or $\phi^*$ in a Schechter function fit, is seen to increase significantly between high- and low-dimensional environments.}
\label{LFmain}
\end{figure}

The resultant conditional luminosity functions for each geometric environment are shown in \fig{LFmain} with jackknife error bars. These conditional LFs reveal a higher number density of luminous galaxies in lower dimensional environments, introducing a vertical shift of the LF. 
 We use a Schechter function (\citejap{Schechter1976}),
 \begin{equation}
\phi(M)=\frac{\ln{10}}{2.5}\phi^{*}10^{0.4(M^{*}-M)(1+\alpha)}\exp{(-10^{0.4(M^{*}-M)})},
\label{sceq}
\end{equation}
 to characterise the magnitude and shape of the luminosity function. Here $\alpha$ describes the power law slope of the faint end, $M^*$ describes the magnitude at which there is a break from the power law, or the `knee' of the LF, and $\phi^*$ describes the normalisation. The solid lines of \fig{LFmain} show best fit Schechter functions for each LF, with the best fit parameters given in \tab{Schtab}. There is a clear increase in the normalisation of the LF from voids to knots, shown by the steady increase of $\phi^*$. The turnover point, $M^*$, of the LFs moves towards brighter magnitudes from voids to knots, suggesting that brighter galaxies have an increased bias towards lower dimensional regions. Note that we expect there to be some environmentally dependent degeneracies in the $\alpha$ and $M^*$ parameters (see Fig. D1 of MNR14).
 
A comparison of the upper and lower panels in \fig{LFmain} shows the impact of the choice of different smoothing scales and thresholds. Using a range of parameters following the optimal black curve in \fig{rmsdnewdots}, it was found that the magnitude of the difference between the conditional LFs increases as the smoothing scale decreases or as the eigenvalue threshold decreases. This tends to introduce only a vertical shift to the functions, characterised by $\phi^*$, whilst the shapes of the luminosity functions do not show significant dependence on the smoothing and threshold parameters. The variation in shape between the LFs of each geometric environment are discussed further in the following section. 

\hspace{-1cm}

\begin{table*}
\begin{tabular}{c c c c c c c r}
\hline
\hline
Environment & $\sigma (\mpcoh)$&$\alpha$  & $\log_{10_{\vphantom{A}}}[{\phi^*}/\,h^{-3}{\rm Mpc^{-3}}]$&  $M^*-5\log{h}$ &$f_{\rm env}$ (\%) & Galaxies (\%) &$\bar{\delta_{8}^{env}}^{\vphantom{A}}$\\ \hline
Voids&4&$-1.25\pm0.02$&$-2.38\pm0.02$&$-20.54\pm0.03^{\vphantom{A^A}}$&$ 59$&18&$-0.16$\\
Sheets&4&$-1.23\pm0.02$&$-1.90\pm0.04$&$-20.83\pm0.04$&$ 29$& 34&$0.81$\\
Filaments&4&$-1.22\pm0.02$&$-1.51\pm0.04$&$-21.01\pm0.05$&$10$&36&$2.38$\\
Knots&4&$-1.27\pm0.06$&$-1.19\pm0.14$&$-21.22\pm0.14_{\vphantom{A_{A}}}$&$ 1 $&12&$4.39$\\
\cdashline{1-8}
Voids&10&$-1.29\pm0.03$&$-2.39\pm0.06$&$-20.69\pm0.06^{\vphantom{A^A}}$&$ 37$&15&$-0.03$\\
Sheets&10&$-1.22\pm0.02$&$-2.05\pm0.03$&$-20.78\pm0.03$ &$39 $& 32&$0.69$\\
Filaments&10&$-1.24\pm0.02$&$-1.75\pm0.03$&$-20.99\pm0.04$&$20$&39&$1.93$\\
Knots&10&$-1.25\pm0.04$&$-1.40\pm0.09$&$-21.07\pm0.09_{\vphantom{A_{A}}}$&$3$ &15&$3.82$\\
\hline
\end{tabular}
\caption{Best-fit parameters found for a non-linear least squares Schechter function (Eq. \ref{sceq}) fit to the conditional LF of each environment, classified with either $(\sigma, \lambda)=(4\mpcoh, 0.4)$ or $(10\mpcoh, 0.1)$. $\alpha$ shows no clear trend with environment, $\phi^*$ shows a significant, steady increase from voids to knots and $M^*$ brightens from voids to knots. Errors are calculated from the standard deviation of the resultant parameters for 9 jackknife realisations. Note that we expect there to be some degeneracy between $\alpha$ and $M^*$. The sixth and seventh columns show the percentage of the volume and the percentage of galaxies within our sample classified as each environment respectively. The final column gives the average local overdensity, $\bar{\delta_{8}^{\rm env}}$, of galaxies in each environment, plotted as the dashed vertical lines in the overdensity histograms of \fig{hists}.}
\label{Schtab}
\end{table*}
 

 \subsection{Reference Schechter functions}

\begin{figure}
	\begin{subfigure}{0.5\textwidth}
		\includegraphics[scale=0.45]{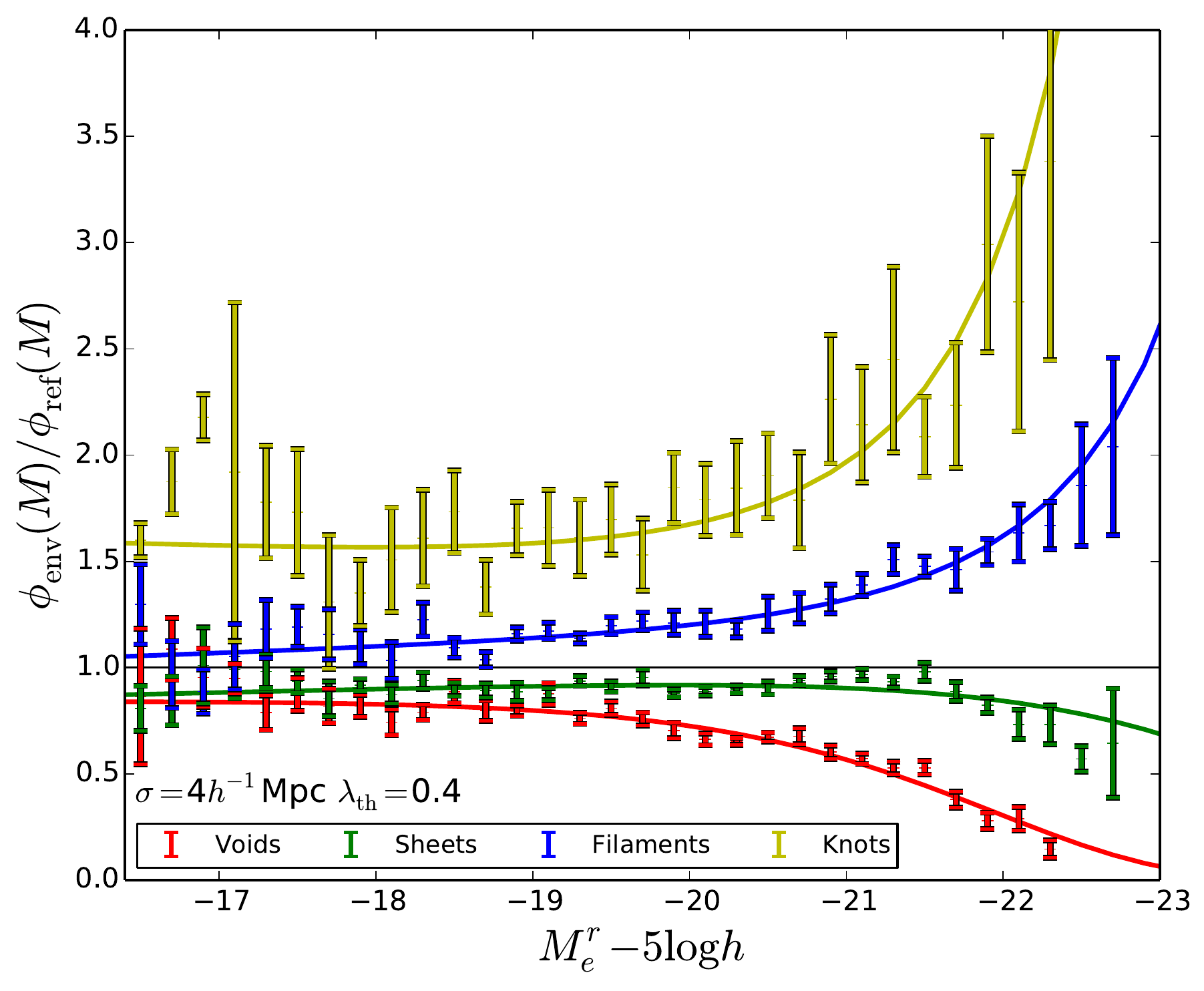}
	 \end{subfigure}

	\begin{subfigure}{0.5\textwidth}
		\includegraphics[scale=0.45]{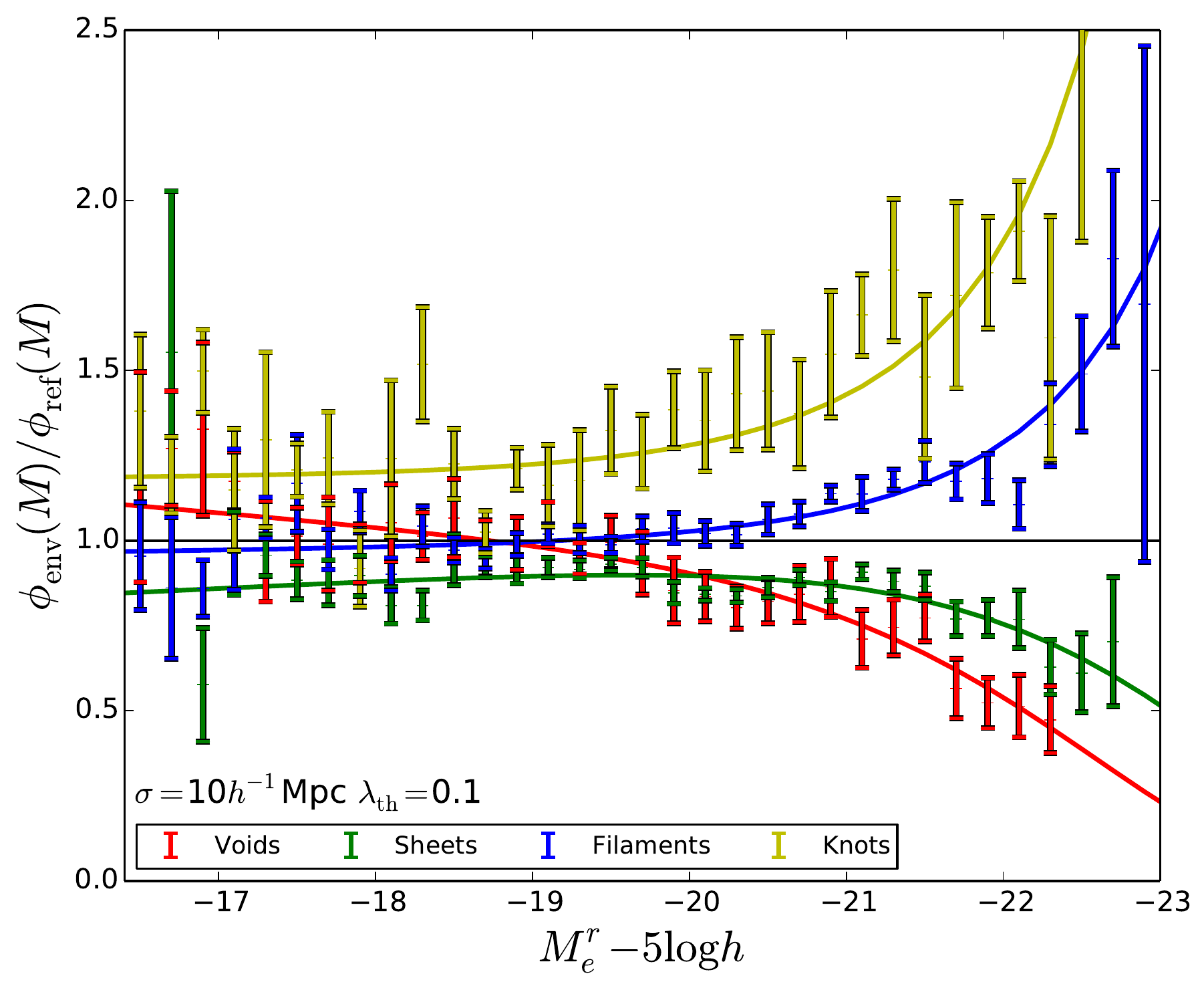}
	\end{subfigure}                    
         \caption{Observed environmental luminosity functions (points) and their best fit Schechter functions (solid) divided by the scaled reference Schechter functions of \eq{refsc} for each geometric environment, colour coded as shown in the legend. The difference in the shape of the LF between the environments is most apparent at the bright end of the LF, owing to the decrease of the turnover magnitude, $M^*$, from voids to knots. Note that the linear scaling means that, for example, a factor of 2 in excess is more noticeable than a factor of 2 in deficit.}                 
  
\label{refschp2}
 \end{figure}

In order to remove some of the vertical offset and clarify the difference in shape between the LFs of each geometric environment, in \fig{refschp2} we plot the ratio of the conditional LFs to a set of scaled reference Schechter functions. The reference function, $\phi_{{\rm ref, tot}}$, is found by fitting a Schechter function to all galaxies in the sample. We apply a normalisation for each environment to produce the scaled reference Schechter functions, $\phi_{{\rm ref, env}}$, given by
 \begin{equation}
\phi_{{\rm ref, env}} = \frac{(1+\bar{\delta}^{\rm env}_{8})}{(1+\bar{\delta}^{\rm tot}_{8})}\times\phi_{{\rm ref, tot}},
\label{refsc}
\end{equation}
where $\bar{\delta}^{\rm{env}}_{8}$ is the average overdensity within an $8\mpcoh$ sphere centred on each galaxy of the environment and $\bar{\delta}^{\rm tot}_{8}$ is that of all the galaxies in the full sample (we find $\bar{\delta}^{\rm tot}_{8}=8\times10^{-3}$). The solid lines in \fig{refschp2} show the ratio of the best fit Schechter functions for each environment to the reference functions, data points show the ratio for the measured luminosity functions. The departure from the global shape is seen to increase towards the bright end of the LF; the number density of void galaxies decreases as we move towards brighter magnitude bins faster than that of the global population whereas we see a slower decline with brightness for the knot galaxies. The remaining vertical offset is likely to be due to the approximations used in defining $\phi_{{\rm ref, env}}$: for example, we do not consider the effect of bias, i.e. galaxies are biased tracers of the underlying dark matter density field and the degree of bias may vary between environments. 

 \subsection{Direct dependence on geometric environment}
 \label{Geodep}
It is clear from the results of MNR14 and others (e.g. \citejap{Hutsi2002}, \citejap{Croton2005}, \citejap{Tempel2011}) that local density plays a significant role in determining the number density of luminous galaxies. In this section we ask whether the geometric environment plays any {\it additional} role. Is it correct to assume that the LF, given a certain local overdensity, will be the same regardless of location within the cosmic web? The analytic results of the dependence of the Schechter fit parameters on local density, $\phi(M|\delta_8)$, presented in MNR14 could be used to answer this question; however, we found statistical uncertainties in the fit parameters at the extremes of $\delta_8$ limited its use. Instead we sample the galaxies in such a way as to remove any additional geometric information from the environment-split subcatalogues, whilst retaining the distribution of local densities. We recalculate the LFs for these resampled catalogues with the hypothesis that any direct modulation by geometric environment will present itself as a disparity between these results.
\begin{figure}
		\includegraphics[scale=0.45]{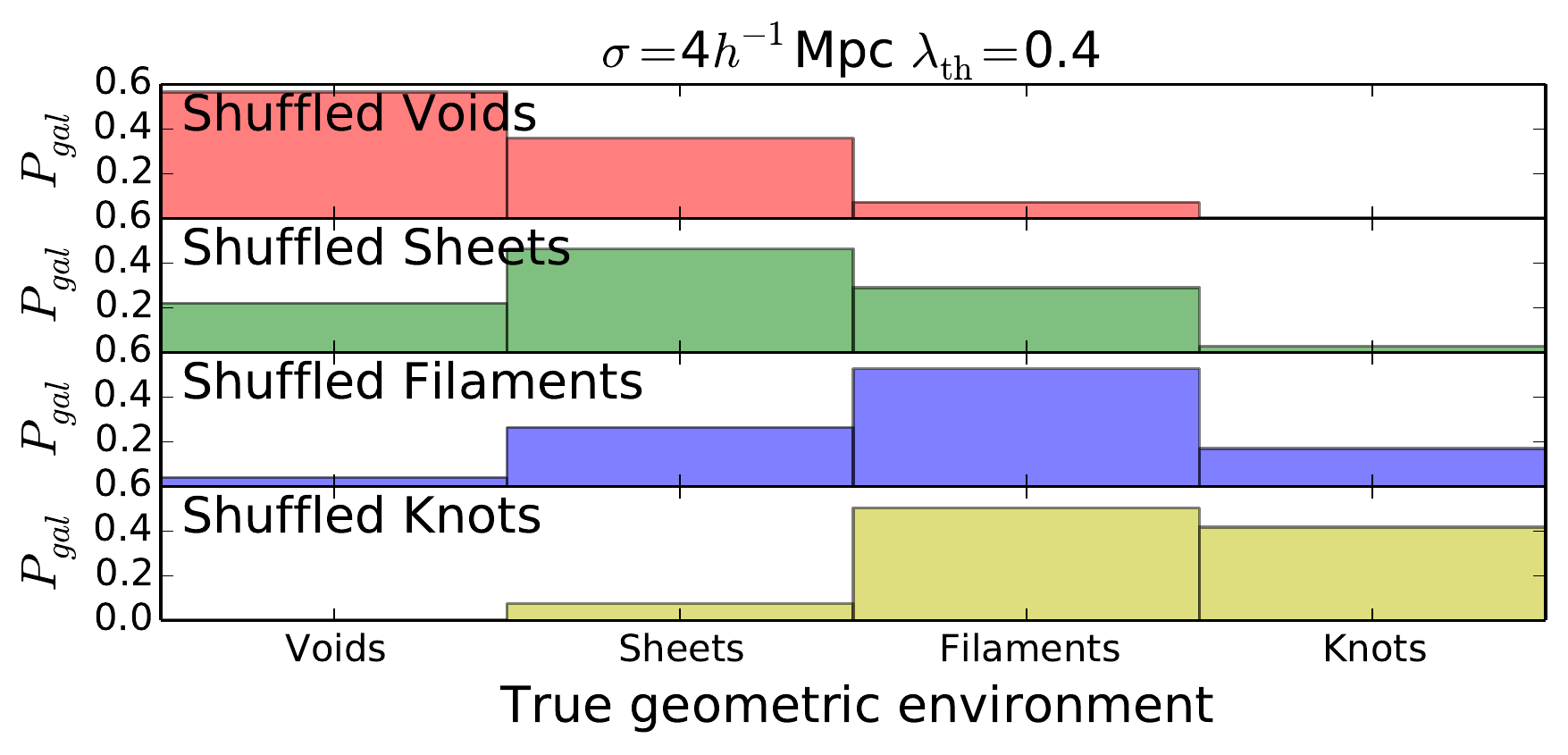} 
	 \caption{The proportion of galaxies from each true geometric environment which were sampled to make up the new shuffled-galaxy catalogues. Shuffled catalogues are predominantly composed of galaxies which were also in the original environment catalogue, as is expected from the distribution of overdensities shown in \fig{hists}. However, there is still significant mixing between environments due to the overlap of the histograms seen in \fig{hists}, which should change the resultant LF if the geometric environment is having a direct influence.}
\label{shufgalhists}
\end{figure}

We populate four `shuffled' catalogues by randomly selecting GAMA galaxies from within overdensity bins, such that the $\delta_{8}$ distribution of each shuffled catalogue matches that of one of the original geometric-environment-split catalogues. This can be thought of as shuffling the galaxies around within regions of the same overdensity (bins of width 0.1 in $\delta_{8}$ are used). For each galaxy that was included in the original LF, we pick at a random a galaxy with the same local overdensity and effectively replace the original galaxy with this new one. Thus the overall effect is to remove the geometric environment distinction contained in the original catalogues, whilst maintaining the same distribution of local densities. 

A volume limited sample is required to allow galaxies to be shuffled randomly over different redshifts without moving galaxies out of their observable redshift range. We use a sample of $\simeq26000$ galaxies satisfying $0.021<z<0.137$ and $-22<M_r^{e}-5\log{h}<-18.5$, chosen as a compromise between a large magnitude range and a large sample size. In \fig{shufgalhists} we show the proportion of galaxies in each shuffled catalogue which were taken from each of the 4 `true' geometric environments defined with our smaller smoothing scale parameter set, $(\sigma,\lambda_{\rm th}) = (4\mpcoh, 0.4)$. For example, the bottom panel of \fig{shufgalhists} shows that $\simeq50\%$ of the galaxies in the shuffled-knots catalogue are from a filament environment, and the remaining $\simeq40\%$ and $\simeq10\%$ of galaxies were drawn from knot and sheet environments respectively. The combined distribution of densities of this selection of galaxies making up the shuffled-knots catalogue matches the distribution seen in the original knots catalogue. A large fraction of the galaxies in each shuffled catalogue were also in the original geometric environment catalogue, as is expected from the distribution of overdensities, but there is still significant mixing due to the overlap of the histograms seen in \fig{hists}. When repeating the shuffling for the $\sigma=10\mpcoh$ geometric environments, we see more mixing between environments due to the slighter broader distribution of densities in any given environment, with more than half the galaxies of each shuffled catalogue being selected from a different geometric environment. We argue that, if geometric environment has a significant direct effect, we should see different luminosity functions for each shuffled catalogue, in which geometric information is lost, as compared with the initial geometrically-split catalogue. 

\begin{figure}
		\includegraphics[scale=0.43]{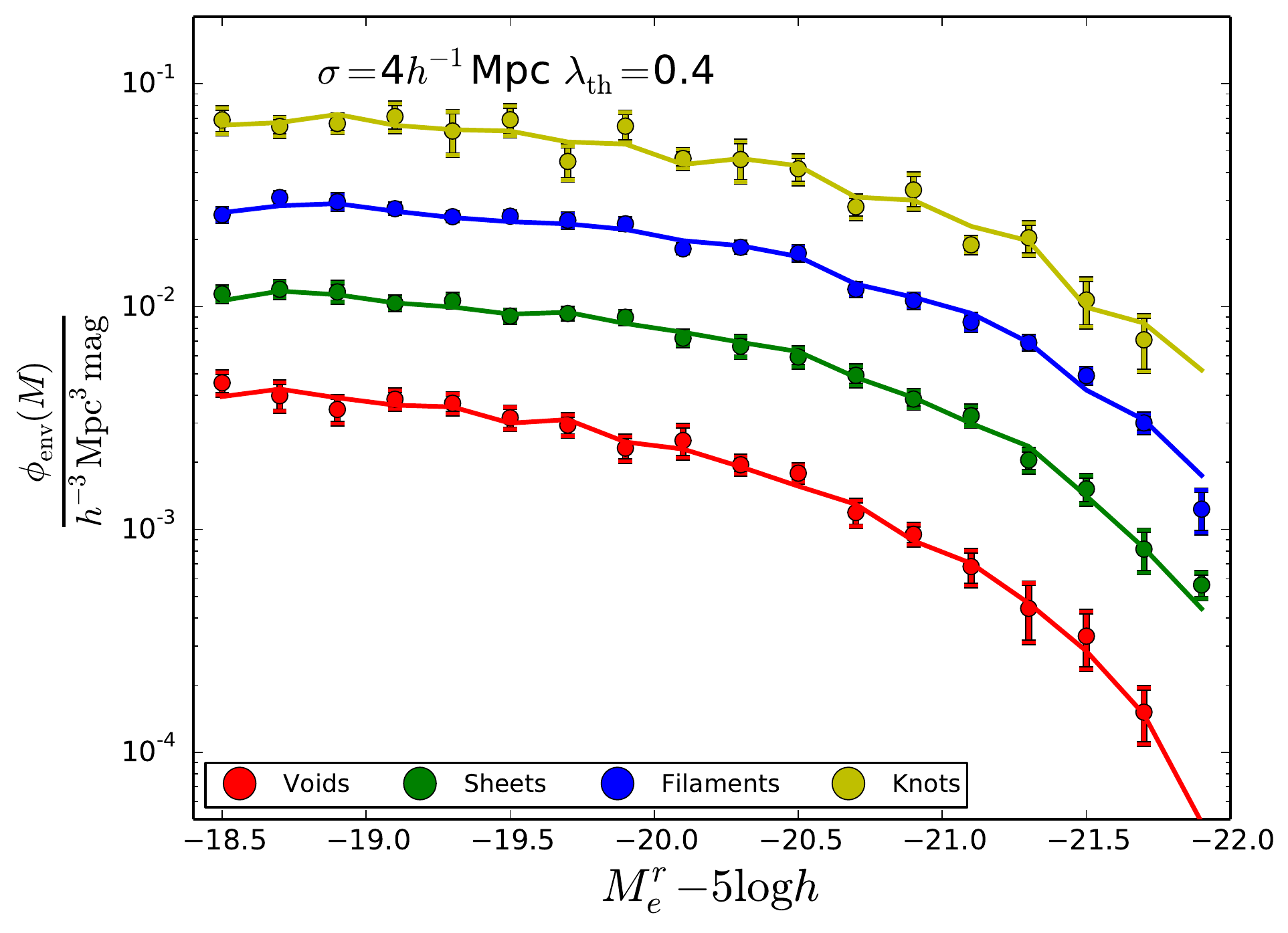}
	\caption{The conditional galaxy LFs of the volume limited sample described in section \ref{Geodep}. The LFs for the true geometric environments are indicated by the circle markers, with jackknife error bars. For each geometric environment we create 9 realisations of a shuffled catalogue which mimics the distribution of overdensities in the geometric environment but selects galaxies randomly regardless of local geometry. The solid lines plot the average of the 9 realisations for each environment and can be seen to be fully consistent with the original LFs, indicating the galaxy LF is independent of geometry for a given overdensity.} 
\label{LFshuf}
\end{figure}

\begin{figure}
		\includegraphics[scale=0.43]{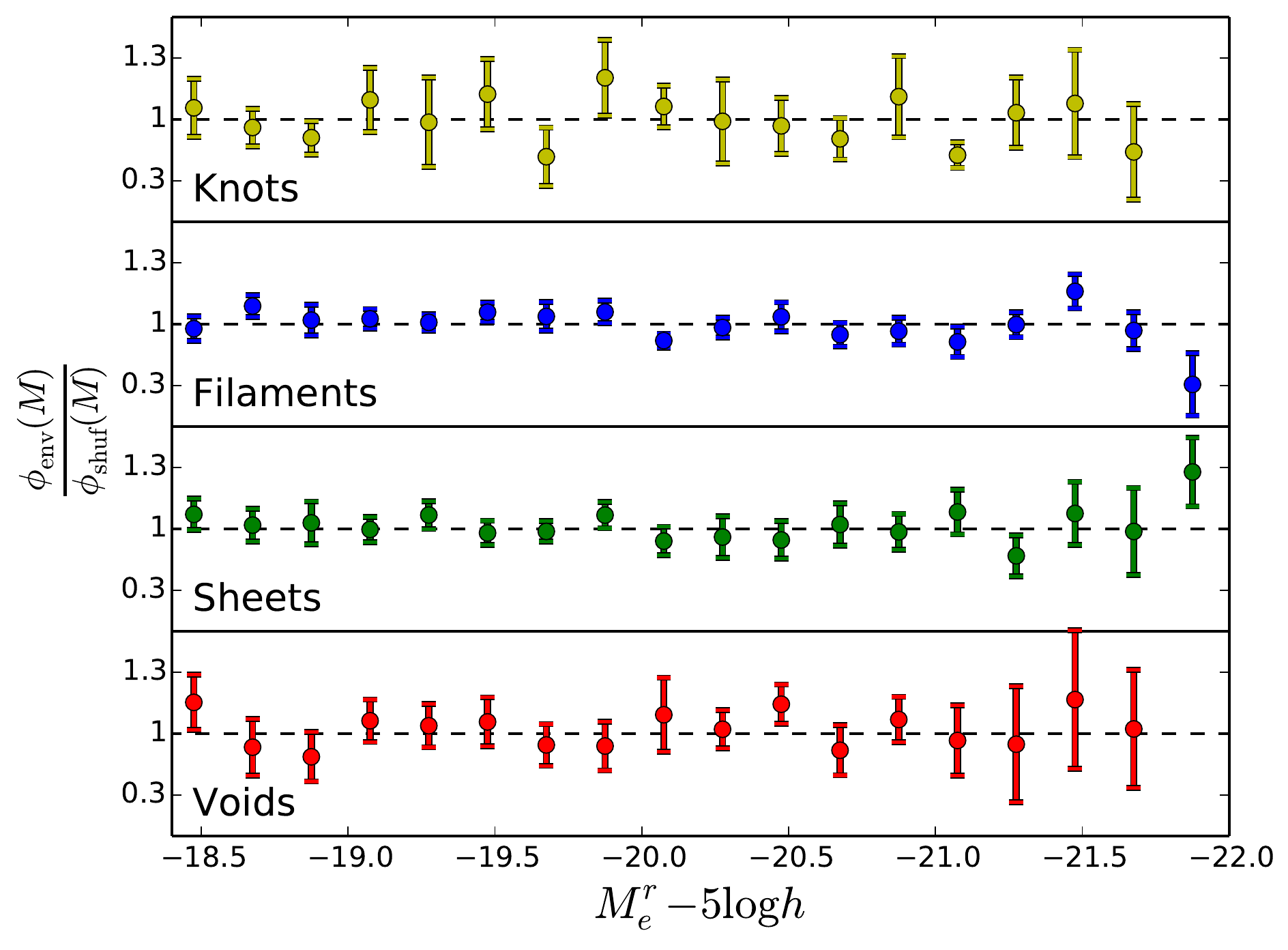}
	\caption{The volume-limited LF and jackknife errors for each geometric environment, divided by the average LF of 9 realisations of shuffled catalogues composed of galaxies selected randomly from the full volume to mimic the density distribution of the corresponding geometric environment. Dashed lines represent a ratio of 1, indicating no variation between the geometric and the shuffled LFs. No statistically significant deviation away from a ratio of 1 is seen, which leads us to conclude the shuffled catalogues are consistent with the original LF and the cosmic web has no detectable direct influence.}
\label{LFshufgalratio}
\end{figure}

\fig{LFshuf} shows the LF for each environment, given by the circles and jackknife error bars, and for an average over 9 realisations of shuffled catalogues, shown by the solid lines. The LFs of the original and the shuffled catalogues are fully consistent, indicating that {\it the local overdensity is the only significant environmental property affecting the galaxy LF and the cosmic web has no direct influence}. The ratio between the geometric- and the shuffled-LFs, shown in \fig{LFshufgalratio}, further emphasises that we find no statistically significant difference between the two measurements. We show only the $(\sigma,\lambda_{\rm th}) = (4\mpcoh, 0.4)$ results here, noting that the same analysis applied to the geometric environments as classified with $(\sigma,\lambda_{\rm th}) = (10\mpcoh, 0.1)$ draws the same conclusion.

We tested the scale dependence of this result by repeating the shuffling process with densities defined over spheres of radii $6$ and $12\mpcoh$, finding no significant differences in our results. As a further test, we shuffled the geometric classifications of individual cells of the initial Cartesian grid within density bins where the density was defined by the smoothed density field (4 or $10\mpcoh$) used to initially generate the geometric classifications. This has the advantage that we do not require a volume limited sample and hence can test the full magnitude range. Again we found no statistically significant difference, reinforcing the main result of this paper, that the galaxy LF is independent of geometry for a given smoothed density.

\section{Summary and discussion}
\label{summary}
In this paper, we have developed a method for probing the cosmic web in galaxy
redshift surveys, which has been applied to the GAMA dataset. Of the many tools
that have been proposed for picking out the skeleton of large-scale structure, we have
rejected those that depend on unobservable velocity information, and focused on
a simple technique based on the tidal tensor -- the Hessian matrix of second derivatives
of the pseudopotential that is sourced by the galaxy density fluctuation. This allows
the survey volume to be dissected into its four distinct geometrical environments --
voids, sheets, filaments and knots -- based
on the number of eigenvalues of the tidal tensor that lie above a threshold. 

We have explored different choices for the
threshold and identified a practical option in which galaxies are distributed
roughly equally between different components of the web, allowing good statistics
in the comparison of properties as a function of environment.
We have carried out tests with simulations to show how this method can yield robust
environmental classifications in the presence of real-world effects such as redshift-space
distortions, bias and survey geometry. In particular, we applied a method where the
data are reflected at the survey boundaries in order to mitigate a bias that would
otherwise arise if the survey was simply zero-padded. This has allowed us to
search for the impact of large-scale tidal forces on the galaxy population, and there have
been a number of studies suggesting that an effect of this sort should be present.

\cite{Metuki2014} conducted a thorough analysis of galaxy properties within the cosmic web, finding significant dependencies on location, which they attributed largely to the strong relationship between galaxy properties and the properties of their host halos, which are in turn linked to their geometric environments. They found that the strong dependence of the halo mass function on the cosmic web was the main cause of the apparent dependence of galaxy properties. However, these analyses often do not test whether the relationships found can be directly attributed to a modulation by geometric environment, or rather are manifestations of the indirect influence of the local density field. In fact, \cite{Alonso2014} show that, based on Gaussian statistics within the linear regime, we expect a variation of the halo mass function within different web components that is due solely to the underlying density field; there is no coupling to tidal forces and the theoretical halo mass function is independent of geometry at a given local density. 

%
\cite{Yan2012} studied the tidal dependence of galaxy properties using the ellipticity, constructed from the eigenvalues of the tidal tensor in a way that exhibits less $\delta$-dependence as a measure of environment than the classification method used in this work. Their analysis of SDSS data revealed no physical influence of environmental morphology on galaxy properties. Similarly, Alpaslan et al. (in prep.) investigated the relations between various galaxy properties and large scale structure identified within the GAMA dataset by the methods discussed in appendix \ref{mehmet}. They found the environment had limited direct impact on galaxy properties, with the stellar mass of a galaxy playing a far larger role in shaping its evolution than the galaxy's location within the cosmic web. However, when identifying filaments by the `Bisous' process, \cite{Guo2014} find a disparity between the LFs of satellite galaxies in SDSS whose host galaxy resides in a filament and those whose host galaxy does not, which they claim cannot be attributed to an environmental bias. Recent work has found a direct relationship between LSS anistropies and the cosmic web when considering tensor properties such as the spin of galaxies (\citejap{Libeskind2012}) and angular momentum of dark matter substructures (\citejap{Dubois2014}). These authors find a correlation between the orientation of LSS and the axes of the cosmic web.

With this context, the main results of the GAMA-based analysis in this paper can be summarised as follows:
\begin{itemize}
\item We have measured the galaxy luminosity function in each component of the web and found a strong variation. By fitting a Schechter function to each conditional luminosity function we quantify the variation of the LF between web components. The normalisation, described by $\phi^*$, increases by a factor of $\sim$10 from voids to knots. The knee of the LF, $M^*$, brightens from voids to knots by $0.7$ and $0.4$ mag for 4 and 10$\mpcoh$ smoothing scales respectively. We find no clear trend in $\alpha$, the parameter describing the faint end slope of the LF. 

\item We test the direct influence of the cosmic web by investigating the extent to which the observed modulation may be attributed to variations in the local density. By measuring densities over a range of scales between $6$ and $12\mpcoh$ we find, in all cases, that the modulation may be entirely accounted for by the variations in density between the geometric environments, indicating that {\it the galaxy LF is independent of geometry at a given local density}.
\end{itemize}

Our results are thus consistent with a picture in which scalar
properties of halo and galaxy populations have no {\it direct\/}
dependence on their location within the geometry of the cosmic web.
Our clearly detected variation in the luminosity function of galaxies
between different geometrical environments can be entirely accounted
for via the known correlation of galaxy properties with local
overdensity, plus the tendency for different locations in the web to
sample different densities.  It would however be premature to argue on
this basis that tidal effects have no impact on the assembly of galaxy
structures. It is possible that such higher-order influences are
simply too small to be detected at the resolution of this
analysis. The scales that we have chosen to probe are set by the
requirement of classifying the entire volume of a survey in a way that
is robust given the limited number density of galaxies. Higher-density
regions could be studied to higher spatial resolution, and it may be
that some of the published claims of detected tidal effects might yet
be validated. But it is clear from the current study that such
effects are highly sub-dominant.

Whilst this paper considers only the galaxy luminosity function, there
are a number of other observable properties that would benefit from a
thorough analysis of the direct influence of the cosmic web. Galaxy
colour, mass, morphology and star formation history may all exhibit
some environmental dependencies. Additionally, there may be reason to
expect satellite, or low-mass galaxies to be more strongly linked to
their geometrical environment (e.g. Carollo et al. 2013). Finally, it
remains to be seen whether claims of anisotropy within different
geometrical environments are intrinsic or whether they can be fully
accounted for by secondary correlations with overdensity, as explored
here for the luminosity function. We can therefore envisage
considerable future applications of the tools we have established for
the practical exploration of the cosmic web in real galaxy surveys.

\section{Acknowledgements}
EE acknowledges support from the Science and Technology Facilities Council.
TMR acknowledges support from a ERC Starting Grant (DEGAS-259586). CH acknowledges support from the ERC under the EC FP7 grant number 240185. PN acknowledges the support of the Royal Society through the award of a University Research Fellowship and the ERC, through receipt of a Starting Grant (DEGAS- 259586). Data used in this paper will be available through the GAMA DB (\url{http://www.gama-survey.org/}) once the associated redshifts are publicly released. GAMA is a joint European-Australian project based around a spectroscopic campaign using the Anglo-Australian Telescope. The GAMA input catalogue is based on data taken from the Sloan Digital Sky Survey and the UKIRT Infrared Deep Sky Survey. Complementary imaging of the GAMA regions is being obtained by a number of independent survey programs including GALEX MIS, VST KIDS, VISTA VIKING, WISE, Herschel-ATLAS, GMRT and ASKAP providing UV to radio coverage. GAMA is funded by the STFC (UK), the ARC (Australia), the AAO, and the participating institutions. The GAMA website is \url{http://www.gama-survey.org/}.

The MultiDark Database used in this paper and the web application providing online access to it were constructed as part of the activities of the German Astrophysical Virtual Observatory as result of a collaboration between the Leibniz-Institute for Astrophysics Potsdam (AIP) and the Spanish MultiDark Consolider Project CSD2009-00064. The Bolshoi and MultiDark simulations were run on the NASA's Pleiades supercomputer at the NASA Ames Research Center. The MultiDark-Planck (MDPL) and the BigMD simulation suite have been performed in the Supermuc supercomputer at LRZ using time granted by PRACE.
  \bibliography{bib_me}

\begin{thebibliography}{}

\bibitem[\protect\citeauthoryear{{Alonso}, {Eardley} \& {Peacock}}{{Alonso}
  et~al.}{2014}]{Alonso2014}
{Alonso} D.,  {Eardley} E.,    {Peacock} J.,  2014, ArXiv: 1406.4159

\bibitem[\protect\citeauthoryear{{Alpaslan} et~al.,}{{Alpaslan}
  et~al.}{2014}]{Alpaslan2014}
{Alpaslan} M.,  et~al., 2014, \mnras, 438, 177

\bibitem[\protect\citeauthoryear{{Arag{\'o}n-Calvo}, {van de Weygaert} \&
  {Jones}}{{Arag{\'o}n-Calvo} et~al.}{2010}]{Aragon-Calvo2010}
{Arag{\'o}n-Calvo} M.~A.,  {van de Weygaert} R.,    {Jones} B.~J.~T.,  2010,
  \mnras, 408, 2163

\bibitem[\protect\citeauthoryear{{Baldry} et~al.,}{{Baldry}
  et~al.}{2010}]{Baldry2010}
{Baldry} I.~K.,  et~al., 2010, \mnras, 404, 86

\bibitem[\protect\citeauthoryear{{Balogh}, {Baldry}, {Nichol}, {Miller},
  {Bower} \& {Glazebrook}}{{Balogh} et~al.}{2004}]{Balogh2004}
{Balogh} M.~L.,  {Baldry} I.~K.,  {Nichol} R.,  {Miller} C.,  {Bower} R.,
  {Glazebrook} K.,  2004, \apj, 615, L101

\bibitem[\protect\citeauthoryear{{Bardeen}, {Bond}, {Kaiser} \&
  {Szalay}}{{Bardeen} et~al.}{1986}]{Bardeen1986}
{Bardeen} J.~M.,  {Bond} J.~R.,  {Kaiser} N.,    {Szalay} A.~S.,  1986, \apj,
  304, 15

\bibitem[\protect\citeauthoryear{{Barrow}, {Bhavsar} \& {Sonoda}}{{Barrow}
  et~al.}{1985}]{Barrow1985}
{Barrow} J.~D.,  {Bhavsar} S.~P.,    {Sonoda} D.~H.,  1985, \mnras, 216, 17

\bibitem[\protect\citeauthoryear{{Brough} et~al.,}{{Brough}
  et~al.}{2013}]{Brough2013}
{Brough} S.,  et~al., 2013, \mnras, 435, 2903

\bibitem[\protect\citeauthoryear{{Cautun}, {van de Weygaert} \&
  {Jones}}{{Cautun} et~al.}{2013}]{Cautun2013}
{Cautun} M.,  {van de Weygaert} R.,    {Jones} B.~J.~T.,  2013, \mnras, 429,
  1286

\bibitem[\protect\citeauthoryear{{Codis}, {Pichon}, {Devriendt}, {Slyz},
  {Pogosyan}, {Dubois} \& {Sousbie}}{{Codis} et~al.}{2012}]{Codis2012}
{Codis} S.,  {Pichon} C.,  {Devriendt} J.,  {Slyz} A.,  {Pogosyan} D.,
  {Dubois} Y.,    {Sousbie} T.,  2012, \mnras, 427, 3320

\bibitem[\protect\citeauthoryear{{Cole}}{{Cole}}{2011}]{Cole2011}
{Cole} S.,  2011, \mnras, 416, 739

\bibitem[\protect\citeauthoryear{{Cole} \& {Kaiser}}{{Cole} \&
  {Kaiser}}{1989}]{Cole1989}
{Cole} S.,  {Kaiser} N.,  1989, \mnras, 237, 1127

\bibitem[\protect\citeauthoryear{{Colless} et~al.,}{{Colless}
  et~al.}{2001}]{Colless2001}
{Colless} M.,  et~al., 2001, \mnras, 328, 1039

\bibitem[\protect\citeauthoryear{{Croton} et~al.,}{{Croton}
  et~al.}{2005}]{Croton2005}
{Croton} D.~J.,  et~al., 2005, \mnras, 356, 1155

\bibitem[\protect\citeauthoryear{{de la Torre} et~al.,}{{de la Torre}
  et~al.}{2013}]{delaTorre2013a}
{de la Torre} S.,  et~al., 2013, \aap, 557, A54

\bibitem[\protect\citeauthoryear{{Dressler}}{{Dressler}}{1980}]{Dressler1980}
{Dressler} A.,  1980, \apj, 236, 351

\bibitem[\protect\citeauthoryear{{Driver} et~al.,}{{Driver}
  et~al.}{2009}]{Driver2009}
{Driver} S.~P.,  et~al., 2009, Astronomy and Geophysics, 50, 12

\bibitem[\protect\citeauthoryear{{Driver} et~al.,}{{Driver}
  et~al.}{2011}]{Driver2011}
{Driver} S.~P.,  et~al., 2011, \mnras, 413, 971

\bibitem[\protect\citeauthoryear{{Dubois} et~al.,}{{Dubois}
  et~al.}{2014}]{Dubois2014}
{Dubois} Y.,  et~al., 2014, \mnras, 444, 1453

\bibitem[\protect\citeauthoryear{{Efstathiou}, {Ellis} \&
  {Peterson}}{{Efstathiou} et~al.}{1988}]{Efstathiou1988}
{Efstathiou} G.,  {Ellis} R.~S.,    {Peterson} B.~A.,  1988, \mnras, 232, 431

\bibitem[\protect\citeauthoryear{{Falck}, {Neyrinck} \& {Szalay}}{{Falck}
  et~al.}{2012}]{Falck2012}
{Falck} B.~L.,  {Neyrinck} M.~C.,    {Szalay} A.~S.,  2012, \apj, 754, 126

\bibitem[\protect\citeauthoryear{{Forero-Romero}, {Contreras} \&
  {Padilla}}{{Forero-Romero} et~al.}{2014}]{Forero-Romero2014}
{Forero-Romero} J.~E.,  {Contreras} S.,    {Padilla} N.,  2014, \mnras, 443,
  1090

\bibitem[\protect\citeauthoryear{{Forero-Romero}, {Hoffman}, {Gottl{\"o}ber},
  {Klypin} \& {Yepes}}{{Forero-Romero} et~al.}{2009}]{Forero-Romero2009}
{Forero-Romero} J.~E.,  {Hoffman} Y.,  {Gottl{\"o}ber} S.,  {Klypin} A.,
  {Yepes} G.,  2009, \mnras, 396, 1815

\bibitem[\protect\citeauthoryear{{Garilli} et~al.,}{{Garilli}
  et~al.}{2014}]{Garilli2014}
{Garilli} B.,  et~al., 2014, \aap, 562, A23

\bibitem[\protect\citeauthoryear{{G{\'o}mez} et~al.,}{{G{\'o}mez}
  et~al.}{2003}]{Gomez2003}
{G{\'o}mez} P.~L.,  et~al., 2003, \apj, 584, 210

\bibitem[\protect\citeauthoryear{{Guo}, {Tempel} \& {Libeskind}}{{Guo}
  et~al.}{2014}]{Guo2014}
{Guo} Q.,  {Tempel} E.,    {Libeskind} N.~I.,  2014, ArXiv: 1403.5563

\bibitem[\protect\citeauthoryear{{Hahn}, {Porciani}, {Carollo} \&
  {Dekel}}{{Hahn} et~al.}{2007}]{Hahn2007}
{Hahn} O.,  {Porciani} C.,  {Carollo} C.~M.,    {Dekel} A.,  2007, \mnras, 375,
  489

\bibitem[\protect\citeauthoryear{{Hahn}, {Porciani}, {Dekel} \&
  {Carollo}}{{Hahn} et~al.}{2009}]{Hahn2009}
{Hahn} O.,  {Porciani} C.,  {Dekel} A.,    {Carollo} C.~M.,  2009, \mnras, 398,
  1742

\bibitem[\protect\citeauthoryear{{Hoffman}, {Metuki}, {Yepes}, {Gottl{\"o}ber},
  {Forero-Romero}, {Libeskind} \& {Knebe}}{{Hoffman}
  et~al.}{2012}]{Hoffman2012}
{Hoffman} Y.,  {Metuki} O.,  {Yepes} G.,  {Gottl{\"o}ber} S.,  {Forero-Romero}
  J.~E.,  {Libeskind} N.~I.,    {Knebe} A.,  2012, \mnras, 425, 2049

\bibitem[\protect\citeauthoryear{{H{\"u}tsi}, {Einasto}, {Tucker}, {Saar},
  {Einasto}, {M{\"u}ller}, {Hein{\"a}m{\"a}ki} \& {Allam}}{{H{\"u}tsi}
  et~al.}{2002}]{Hutsi2002}
{H{\"u}tsi} G.,  {Einasto} J.,  {Tucker} D.~L.,  {Saar} E.,  {Einasto} M.,
  {M{\"u}ller} V.,  {Hein{\"a}m{\"a}ki} P.,    {Allam} S.~S.,  2002,
  arXiv:astro-ph/0212327

\bibitem[\protect\citeauthoryear{{Kauffmann}, {White} \&
  {Guiderdoni}}{{Kauffmann} et~al.}{1993}]{Kauffmann1993}
{Kauffmann} G.,  {White} S.~D.~M.,    {Guiderdoni} B.,  1993, \mnras, 264, 201

\bibitem[\protect\citeauthoryear{{Libeskind}, {Hoffman}, {Knebe}, {Steinmetz},
  {Gottl{\"o}ber}, {Metuki} \& {Yepes}}{{Libeskind}
  et~al.}{2012}]{Libeskind2012}
{Libeskind} N.~I.,  {Hoffman} Y.,  {Knebe} A.,  {Steinmetz} M.,
  {Gottl{\"o}ber} S.,  {Metuki} O.,    {Yepes} G.,  2012, \mnras, 421, L137

\bibitem[\protect\citeauthoryear{{Ludlow} \& {Porciani}}{{Ludlow} \&
  {Porciani}}{2011}]{Ludlow2011}
{Ludlow} A.~D.,  {Porciani} C.,  2011, \mnras, 413, 1961

\bibitem[\protect\citeauthoryear{{McNaught-Roberts} et~al.,}{{McNaught-Roberts}
   et~al.}{2014}]{Tam14}
{McNaught-Roberts} T.,  et~al., 2014, Astrophysics

\bibitem[\protect\citeauthoryear{{Metuki}, {Libeskind}, {Hoffman}, {Crain} \&
  {Theuns}}{{Metuki} et~al.}{2014}]{Metuki2014}
{Metuki} O.,  {Libeskind} N.~I.,  {Hoffman} Y.,  {Crain} R.~A.,    {Theuns} T.,
   2014, ArXiv: 1405.0281

\bibitem[\protect\citeauthoryear{{Moster}, {Somerville}, {Maulbetsch}, {van den
  Bosch}, {Macci{\`o}}, {Naab} \& {Oser}}{{Moster} et~al.}{2010}]{Moster2010}
{Moster} B.~P.,  {Somerville} R.~S.,  {Maulbetsch} C.,  {van den Bosch} F.~C.,
  {Macci{\`o}} A.~V.,  {Naab} T.,    {Oser} L.,  2010, \apj, 710, 903

\bibitem[\protect\citeauthoryear{{Platen} et~al.,}{{Platen}
  et~al.}{2007}]{Platen2007}
{Platen} E.,  et~al., 2007, \mnras, 380, 551

\bibitem[\protect\citeauthoryear{{Prada}, {Klypin}, {Cuesta}, {Betancort-Rijo}
  \& {Primack}}{{Prada} et~al.}{2012}]{Prada2012}
{Prada} F.,  {Klypin} A.~A.,  {Cuesta} A.~J.,  {Betancort-Rijo} J.~E.,
  {Primack} J.,  2012, \mnras, 423, 3018

\bibitem[\protect\citeauthoryear{{Robotham} et~al.,}{{Robotham}
  et~al.}{2010}]{Robotham2010b}
{Robotham} A.,  et~al., 2010, PASA, 27, 76

\bibitem[\protect\citeauthoryear{{Robotham} et~al.,}{{Robotham}
  et~al.}{2013}]{Robotham2013}
{Robotham} A.~S.~G.,  et~al., 2013, \mnras, 431, 167

\bibitem[\protect\citeauthoryear{{Schechter}}{{Schechter}}{1976}]{Schechter1976}
{Schechter} P.,  1976, \apj, 203, 297

\bibitem[\protect\citeauthoryear{Sheth, Mo \& Tormen}{Sheth
  et~al.}{2001}]{Sheth2001}
Sheth R.~K.,  Mo H.~J.,    Tormen G.,  2001, \mnras, 323, 1

\bibitem[\protect\citeauthoryear{{Sousbie}}{{Sousbie}}{2011}]{Sousbie2011}
{Sousbie} T.,  2011, \mnras, 414, 350

\bibitem[\protect\citeauthoryear{{Tempel}, {Saar}, {Liivam{\"a}gi}, {Tamm},
  {Einasto}, {Einasto} \& {M{\"u}ller}}{{Tempel} et~al.}{2011}]{Tempel2011}
{Tempel} E.,  {Saar} E.,  {Liivam{\"a}gi} L.~J.,  {Tamm} A.,  {Einasto} J.,
  {Einasto} M.,    {M{\"u}ller} V.,  2011, \aap, 529, A53

\bibitem[\protect\citeauthoryear{{Vale} \& {Ostriker}}{{Vale} \&
  {Ostriker}}{2004}]{Vale2004}
{Vale} A.,  {Ostriker} J.~P.,  2004, \mnras, 353, 189

\bibitem[\protect\citeauthoryear{{van den Bosch}, {Yang}, {Mo}, {Weinmann},
  {Macci{\`o}}, {More}, {Cacciato}, {Skibba} \& {Kang}}{{van den Bosch}
  et~al.}{2007}]{vdBosch2007}
{van den Bosch} F.~C.,  {Yang} X.,  {Mo} H.~J.,  {Weinmann} S.~M.,
  {Macci{\`o}} A.~V.,  {More} S.,  {Cacciato} M.,  {Skibba} R.,    {Kang} X.,
  2007, \mnras, 376, 841

\bibitem[\protect\citeauthoryear{{Wijesinghe} et~al.,}{{Wijesinghe}
  et~al.}{2012}]{Wijesinghe2012}
{Wijesinghe} D.~B.,  et~al., 2012, \mnras, 423, 3679

\bibitem[\protect\citeauthoryear{{Yan}, {Fan} \& {White}}{{Yan}
  et~al.}{2012}]{Yan2012}
{Yan} H.,  {Fan} Z.,    {White} S.~D.~M.,  2012, ArXiv: 1203.1225

\bibitem[\protect\citeauthoryear{{York} et~al.,}{{York}
  et~al.}{2000}]{York2000}
{York} D.~G.,  et~al., 2000, {The Astronomical Journal}, 120, 1579

\bibitem[\protect\citeauthoryear{{Zel'dovich}}{{Zel'dovich}}{1970}]{Zeldovich1970}
{Zel'dovich} Y.~B.,  1970, AAP, 5, 84

\end{thebibliography}
 \begin{appendix}

\section{Effects of the survey geometry}
\label{ee}
A significant limitation of the tidal tensor prescription, or of any analysis requiring information on the smoothed density field, is that observational datasets lack information beyond the surveyed region. There is then the question of how to smooth the galaxy distribution near the survey boundaries. If one `zero-pads' the volume outside of the survey, by setting it all to the large-scale average overdensity ($\bar{\delta}=0$ by definition) this will, on average, reduce the magnitude of both overdensities and underdensities that may be straddling the border of the survey and alter our estimate of the true underlying density in a systematic but unpredictable way. To mitigate this effect, we `reflect' the galaxies along each field's boundaries in right ascension ($ra$) and declination ($dec$). We clone the galaxies inside the survey volume and give them an appropriate reflected location outside of the survey volume before the smoothing process. In effect, we make the reasonable assumption that large-scale features continue smoothly beyond the survey edge.

\begin{figure}
 \includegraphics[scale=0.45]{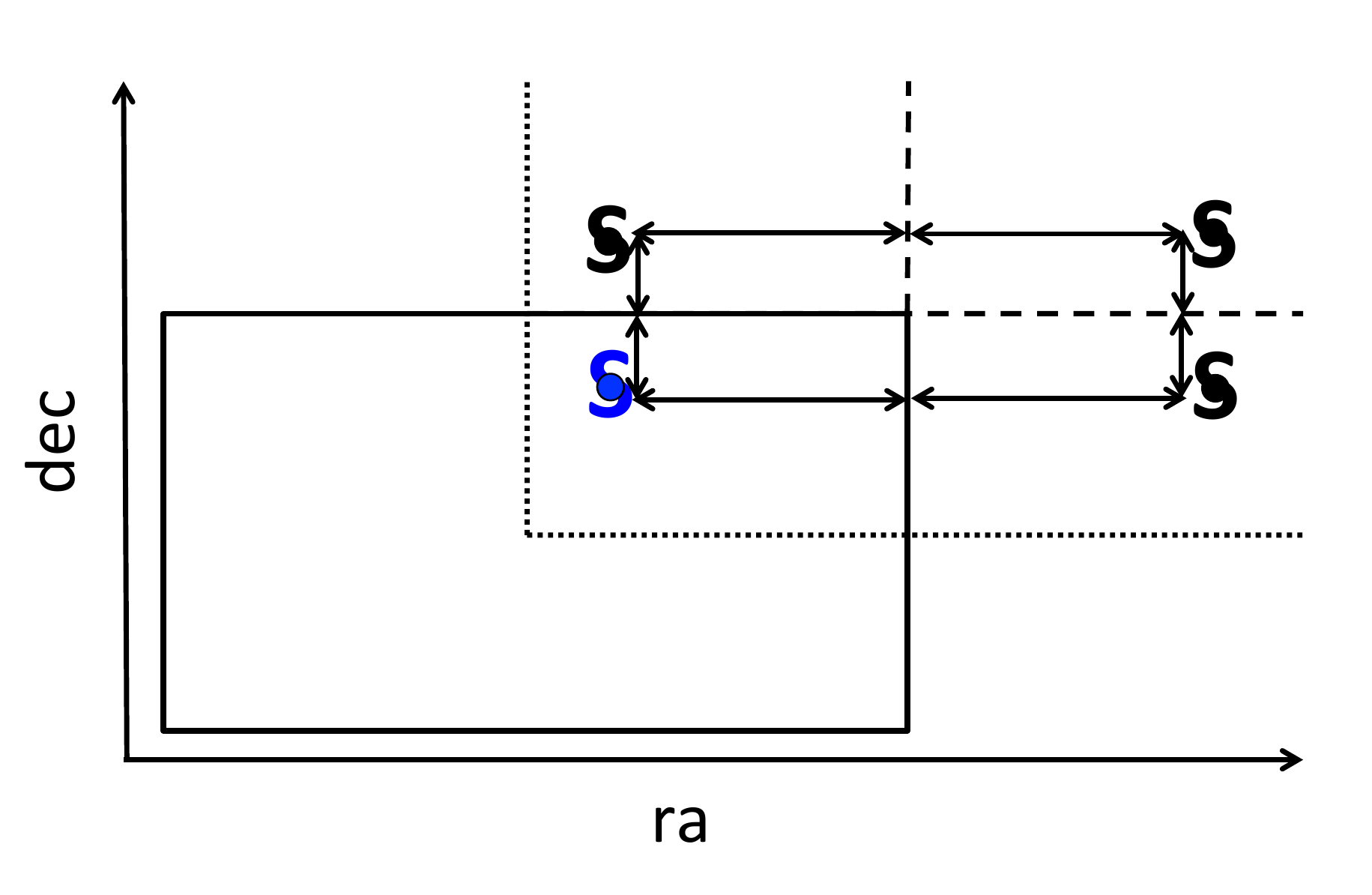}
    	\caption{A sketch of the reflection method. Each galaxy in the survey sample, represented by the blue spiral, is cloned 3 times as depicted by the black spirals. The solid line box represents the ra and dec boundaries of the GAMA field, with redshift pointing out of the page. The dotted lines show the quadrant which is reflected along the nearest two boundaries of the field, which are of constant ra or dec and shown by the dashed lines. Each long arrow in the figure represents the same difference in right ascension, similarly each short arrow represents the same difference in declination.}
    \label{refsketch}
\end{figure}

\begin{figure}
\makebox[0.5\linewidth][l]
    {
    \begin{subfigure}{0.2\textwidth}
    \includegraphics[scale=0.9]{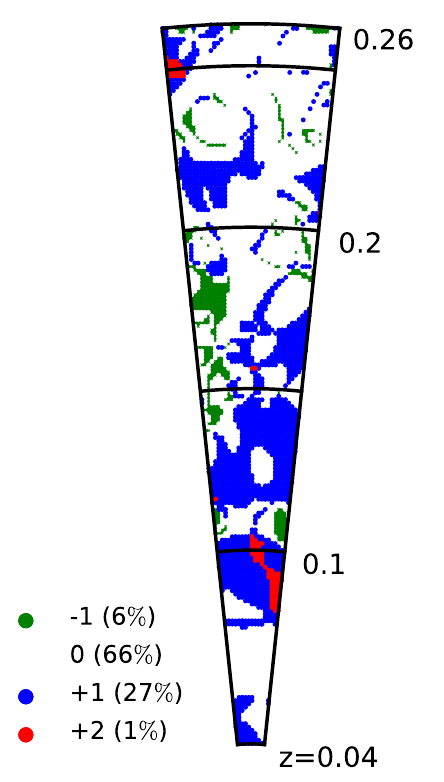}
    \caption{Zero-padding}
    	\label{fig:galdist}
	\end{subfigure}
	
	 \begin{subfigure}{0.2\textwidth}
	 \includegraphics[scale=0.9]{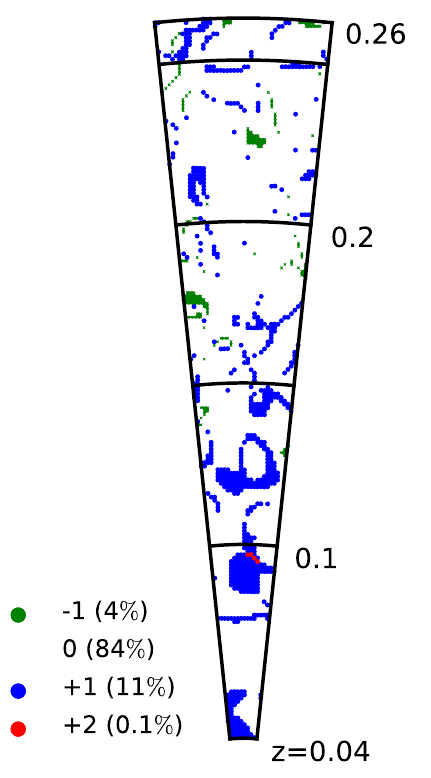}
	 
    \caption{Reflected galaxies}
    	\label{fig:galenv1}
	\end{subfigure}
    }
    	\caption{A test of the effect of the survey geometry on the resulting geometric environment classifications using simulated data. Coloured regions of the figure show the cells which are classified differently to the full simulation results when regions outside a GAMA sized survey cone are zero-padded ({\bf left} panel) or filled with reflected galaxies ({\bf right} panel), as described in the text. This is for an example realisation of a GAMA field and ($\sigma$, $\lambda_{\rm th}$)=(10$\mpcoh$, 0.1). Colour code in the keys refer to the difference, $\Delta{N}$, in the number of eigenvalues above the threshold, $N^{\rm+}$, between the full simulation and the limited-information survey classifications, e.g. $\Delta{N}=N^{\rm+}_{\rm FULL}-N^{\rm+}_{\rm 0pad}$. Hence each cell has a discrete value of $\Delta{N}$, with  $-3\leq\Delta{N}\leq3$. The percentage of all cells with a given $\Delta{N}$, measured over three realisations, is indicated in the keys. We wish to maximise the percentage with $\Delta{N}=0$, as this indicates the limited information has not changed the resultant environment classification of these cells. The increase in $\Delta{N}=0$ from $66\%$ to $84\%$ shows that the reflection technique offers a strong improvement over zero-padding alone.}
    \label{edgeref}
\end{figure}

A sketch illustrating the reflection process is shown in \fig{refsketch}. Each galaxy is cloned 3 times always keeping its original redshift, the first clone is given a new $ra$ equivalent to a reflection along the nearest $ra$ border of the field, the second clone keeps the $ra$ of the original galaxy but has its $dec$ changed to simulate a reflection along the nearest $dec$ boundary of the field, and the third takes on both of these two new $ra$ and $dec$ values. This has an approximately equivalent effect to using a weighted smoothing kernel, where cells near the edge of the volume are up-weighted to account for the lack of information in cells outside of the volume. However, the reflection approach permits the smoothing operation to be a single convolution, giving us the speed advantage of the fast Fourier transform. Beyond the reflection regions (half the width of the survey dimension) we use zero-padding. In the redshift direction, for each field we make use of the full GAMA galaxy catalogue, $0.0 < z < 0.5$, and calculate a density contrast for the full surveyed volume of each field, though, as described in the text, we select only galaxies satisfying $0.04<z<0.263$ for our scientific analyses. 
We make use of the MultiDark (1$\,h^{-1}$Gpc)$^{3}$ dark matter simulation (\citejap{Prada2012}), populated with mock galaxies using halo occupation distribution modelling (\citejap{delaTorre2013a}), selecting GAMA-representative regions from the full simulation where necessary to test this reflection method. The simulated dataset we use is a single redshift snapshot of $z=0.1$ with galaxies randomly sampled so that the number density of mock galaxies matches the average number density of galaxies in our GAMA sample. \fig{edgeref} shows, for the simulated data, the regions in which the resulting environments differ from those when information of the full periodic cube is used when only the GAMA-sized volume information is kept, and other regions are either zero-padded only, or populated with the cloned galaxies as described above. We show here results for the $10\mpcoh$ smoothing, noting that the $4\mpcoh$ smoothing is less affected by the survey geometry (due a reduced `skin-depth' of volume which is significantly affected by the volume outside of the survey), but shows a similar improvement when using this reflection technique. The differences are not confined to the edges of the survey due to the use of Fourier transforms but instead tend to occur along boundaries between regions of different environments due to a slight change in the calculated eigenvalues. The percentages of cells classified differently, measured over three realisations, are given in the key of \fig{edgeref}. It can be seen that the reflection technique is beneficial, increasing the percentage of correctly classified cells from $66\%$ to $84\%$ so that the classifications more closely mimic the results of the full simulation than when zero-padding alone is used. There are remaining unavoidable discrepancies due to the lack of information, however we note that $99.9\%$ of cells are classified within $\pm1$ dimension of environment from the `full-information' results when the reflection technique is applied, an increase from the corresponding value for zero-padding of $99\%$. 

\section{Redshift space distortions and other complications}
\label{rsdsec}
\begin{figure}
\makebox[0.5\linewidth][l]
{
    \begin{subfigure}{0.2\textwidth}
\includegraphics[scale=0.9]{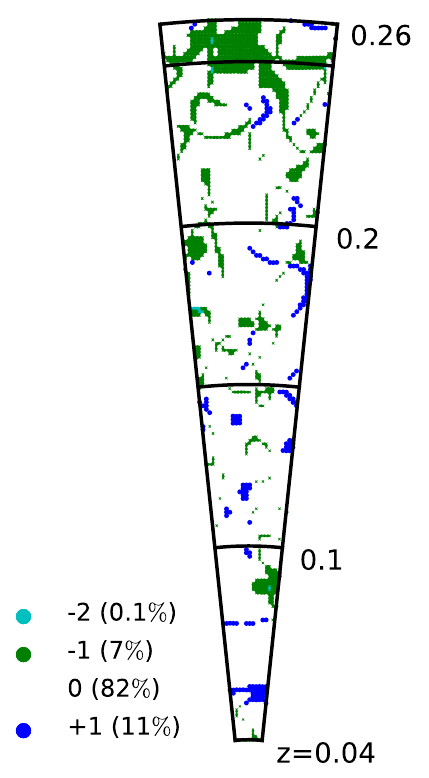}
    \caption{DM vs galaxy $\delta$}
\label{DMdifa}

	\end{subfigure}
	    \begin{subfigure}{0.2\textwidth}
\includegraphics[scale=0.9]{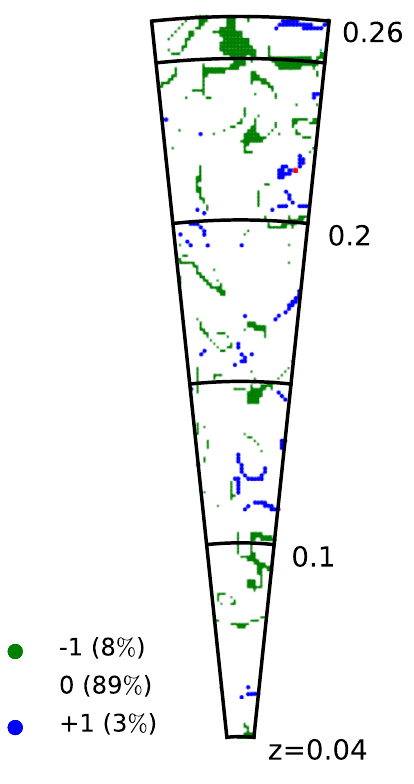}
    \caption{Real- vs Redshift-space}
	 \label{RSdifa}

	\end{subfigure}

}
	\caption{A test of the effects of redshift-space distortions and of using a density field estimated from galaxy number counts on the resulting geometric environment classifications within simulated data. Coloured regions of the figure show the cells which are classified differently, with ($\sigma$, $\lambda$)=(10$\mpcoh$, 0.1), when the dark matter density field is used or the density field is calculated from the (real-space) galaxies ({\bf left} panel) and when the density field is calculated from the real-space galaxies or from redshift-space galaxies ({\bf right} panel). Colour code in the keys refer to the difference in the number of eigenvalues above the threshold, $N^{+}$, between the full simulation and the limited-information survey-style classifications, eg. $N^{+}_{\rm DM}-N^{+}_{\rm gal}$ or $N^{+}_{\rm real-sp}-N^{+}_{\rm redshift-sp}$. The percentages of cells with each difference value, measured over three realisations, are indicated in the keys.}

	\label{DMdif}
\end{figure}

As discussed in section \ref{NI}, the use of biased galaxies in redshift space to estimate the underlying matter overdensity requires caution. We again make use of the simulated dataset to investigate the magnitude of these effects on the resulting environment classifications. The MultiDark simulation provides information on both the underlying dark matter density field and a simulated galaxy catalogue with galaxy velocity information. This allows us to see how the resulting classifications vary when the density field is estimated from the locations of galaxies and when the underlying dark matter density field is used directly. In a similar manner to \fig{edgeref}, \fig{DMdif}a shows those cells which change their classification when the galaxy density field rather than the dark matter density field is used. The use of galaxies to estimate the density results in $20\%$ of the volume appearing to belong to a different geometric environment. 

With the velocity information we are able to shift each galaxy in the simulation to its redshift-space coordinates, by estimating the distance which would have been inferred given its location and radial velocity, and again compute this comparison. \fig{DMdif}b compares the classifications for density fields constructed from redshift- and real-space galaxies. We find the redshift-space distortions to have no effect on $90\%$ of the volume for both 4 and 10$\mpcoh$ smoothing scales.  

\begin{figure}
\centering
       \includegraphics[scale=0.9]{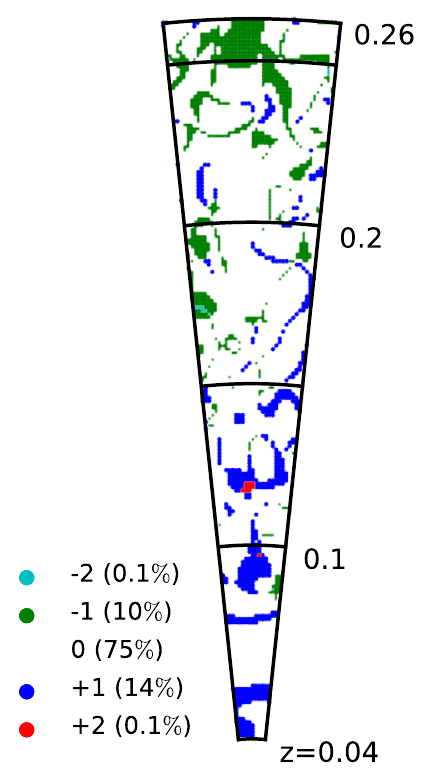}

 
    	\caption{A test of the effects of limited information in observational catalogues on resulting geometric environment classifications. Full-information results are computed from the underlying DM density field using the full periodic 1$h^{-1}$Gpc simulation. Limited-information results use galaxies from the simulation, in redshift space, discarding information outside of a volume representative of a GAMA field and implementing the reflection technique described in the text. Coloured regions of the figure show the cells which are classified differently between the two approaches for an example realisation of a GAMA field and ($\sigma$, $\lambda_{\rm th}$)=(10$\mpcoh$, 0.1). Colour code in the key refers to the difference, $\Delta{N}$, in the number of eigenvalues above the threshold, $N^{\rm+}$, between the full simulation and the limited-information survey classifications, e.g. $\Delta{N}=N^{\rm+}_{\rm FULL}-N^{\rm+}_{\rm LIM}$. Hence each cell has a discrete value of $\Delta{N}$, with  $-3\leq\Delta{N}\leq3$. The percentage of all cells with a given $\Delta{N}$,  within three realisations, is indicated in the key.}
    \label{dmfull}
\end{figure}

We find the combined effect of the three main causes of error when applying the tidal tensor prescription to observational data (survey geometry, a density field sourced from the galaxy number density and redshift-space distortions) to be a change in the resulting geometric environment of $<25\%$ of the volume for both 4 and 10$\mpcoh$ smoothing scales. A example realisation of a field is shown in \fig{dmfull}, indicating the regions which are classified differently when the three causes of error discussed above are all introduced. 

\section{Other GAMA LSS analyses}
\label{mehmet}
\begin{figure}
\includegraphics[width=0.4\textwidth]{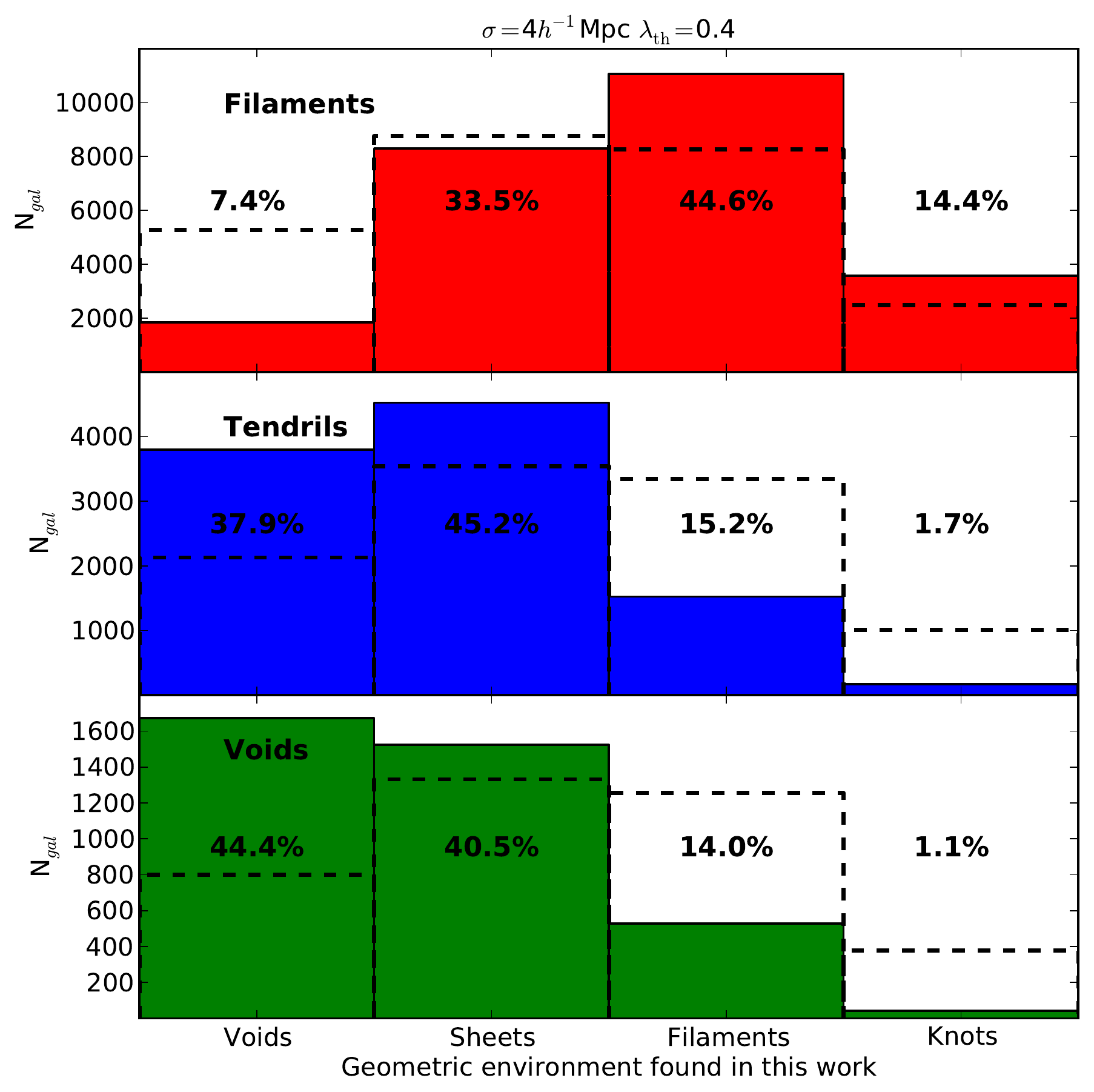}
\caption{Comparison between large scale structure identified in A14 and in this work. Each row shows how either the filaments, tendrils or voids identified in A14 are classified in this work, with $\lambda_{\rm th}=0.4$ and $\sigma=4\mpcoh$. The percentages given in the figure show, for each A14 population, the percentage of galaxies in each of the environments in this work. Dashed lines indicate the number of galaxies in the full GAMA sample classified in each of our geometric environments, normalised by the number of galaxies in the A14 population which each row represents. Hence the dashed lines, which are the same for each panel before normalisation, can be thought of as the expected distribution of a random selection from all galaxies.}
\label{histmem1}
\end{figure}

\begin{figure*}
\makebox[1.2\linewidth][c]{%
\hspace{-1.5cm}

    \begin{subfigure}{0.2\textwidth}
        \includegraphics[scale=1.26]{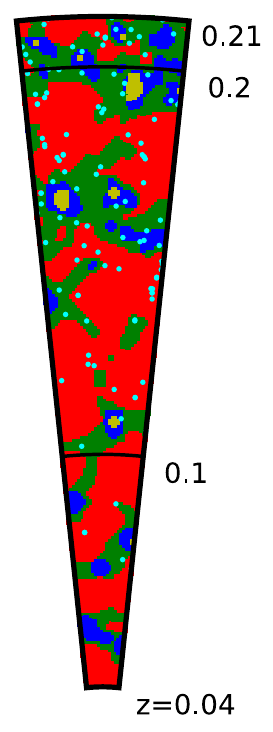}
        \caption{A14 Voids}
   \end{subfigure}
   \hspace{-1.6cm}

    \begin{subfigure}{0.05\textwidth}
   \vspace{5cm}
        \includegraphics[scale=1.1]{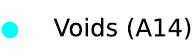}
   \end{subfigure}
\hspace{1.7cm}
   \hspace{1.2cm}
    \begin{subfigure}{0.2\textwidth}
\includegraphics[scale=1.26]{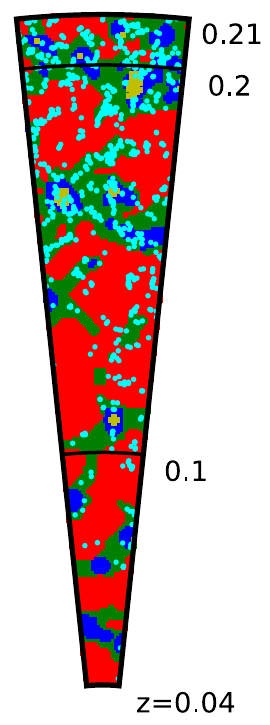}
\caption{A14 Tendrils}
\end{subfigure}
   \hspace{-1.6cm}

    \begin{subfigure}{0.05\textwidth}
   \vspace{5cm}
        \includegraphics[scale=1.1]{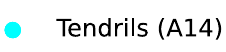}
   \end{subfigure}
      \hspace{1.3cm}
         \hspace{1.6cm}
    \begin{subfigure}{0.2\textwidth}
  \hspace{1.7cm}
        \includegraphics[scale=1.26]{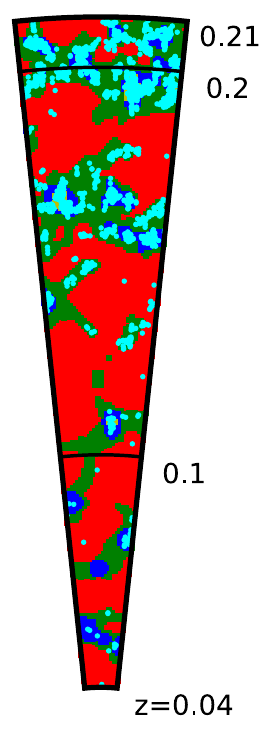}
       \caption{A14 Filaments}
    \end{subfigure}
       \hspace{-1.6cm}

    \begin{subfigure}{0.05\textwidth}
   \vspace{5cm}
        \includegraphics[scale=1.1]{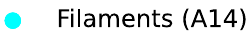}
   \end{subfigure}
     \begin{subfigure}{0.2\textwidth}
        \includegraphics[scale=1.]{ENVcolourkey.pdf}
    \end{subfigure}
 
 }%
    \caption{A comparison of large scale structure identified by this work, and by A14, within the central declination of the G9 field. Our geometric environments, calculated with $\lambda_{\rm th}=0.4$ and $\sigma=4\mpcoh$, are shown by the background colours with red, green, blue and yellow indicating voids, sheets, filaments and knots respectively. From left to right the cyan dots in the figures show the positions of all galaxies within $\pm0.5\degree$ of the central declination in the A14 populations voids, tendrils and filaments respectively.}
    \label{fig:4cones}
\end{figure*}

A previous analysis of large scale structure within the GAMA regions was conducted by \citejap{Alpaslan2014} (hereafter A14). A14 implemented a minimal spanning tree algorithm to identify 643 filaments within the same three GAMA equatorial regions used in this work, with a slightly lower redshift cut of $z<0.213$. A14 also identified a secondary population of smaller coherent structures, tendrils, and a population of isolated void galaxies. In \fig{fig:4cones} we plot the central declination slice of the G9 field, the geometric environments as classified by this work, and all objects in each of the three populations as identified in A14 within $\pm0.5\degree$ of the central declination. We find the filaments of A14 to be visually consistent with the filamentary regions identified in this work. The tendrils and voids of A14 favour the underdense environments of voids and sheets. Note that we show here results for our environments computed with $\sigma=4\mpcoh$, a similar scale to the $r=4.13\mpcoh$ used in A14 as the maximum distance allowed between a galaxy and a filament. We suggest that the `void galaxies' as identified in A14 should be thought of as isolated galaxies, whereas our voids correspond to larger geometric structures. A more quantitative comparison is presented in \fig{histmem1}; the histograms illustrate, for each A14 population, the number of galaxies belonging to each of our geometric environments. The dashed lines in the figure indicate the number of galaxies in the full GAMA sample in each of our environments, normalized by the size of the each A14 population, hence the dashed lines indicate the proportion which would be expected from a purely random selection. 


\clearpage

\end{appendix}

\end{document}